\documentclass{article}
\pdfoutput=1
\usepackage{jheppub}

\usepackage{amsmath}
\usepackage{ulem}
\usepackage{float}

\newcommand{\roughly}[1]{\mathrel{\raise.3ex\hbox{$#1$\kern-0.85em
\lower1ex\hbox{$\sim$}}}}

\newcommand{\lsim}{\roughly<}

\def\exd{{\hbox{d}}}

\def\bea{\begin{eqnarray}}
\def\eea{\end{eqnarray}}
\def\be{\begin{equation}}
\def\ee{\end{equation}}

\def\tr{{\rm tr}\,}

\def\bfx{{\bf x}}
\def\bfy{{\bf y}}

\def\ssI{{\scriptscriptstyle I}}

\def\ssM{{\scriptscriptstyle M}}

\def\cH{\mathcal{H}}

\def\cO{\mathcal{O}}

\def\cR{\mathcal{R}}
\def\cS{\mathcal{S}}

\def\cW{\mathcal{W}}

\def\eff{{\rm eff}}

\def\nn{\nonumber}

\def\({\left(}
\def\){\right)}

\def\pref#1{(\ref{#1})}

\usepackage{braket}
\usepackage{pifont}
\newcommand{\cmark}{\text{\ding{51}}}%
\newcommand{\xmark}{\text{\ding{55}}}
\newcommand{\bR}{ {\mathbb{R}} }
\newcommand{\bC}{ {\mathbb{C}} }
\newcommand{\bx}{ {\mathbf{x}} }

\newcommand{\CO}{ {\mathcal{C}_{\Omega}} }

\newcommand{\DO}{ {\Delta_{\Omega}} }

\newcommand{\vacBD}{ {{\mathrm{BD}}} }
\newcommand{\WBD}{ {\mathcal{W}_{\mathrm{BD}}} }
\newcommand{\RBD}{ {\mathcal{R}_{\mathrm{BD}}} }
\newcommand{\SBD}{ {\mathcal{S}_{\mathrm{BD}}} }
\newcommand{\CBD}{ {\mathcal{C}_{\mathrm{BD}}} }
\newcommand{\CBDp}{ {\mathcal{C}_{\mathrm{BD}}^{\prime}} }
\newcommand{\DBD}{ {\Delta_{\mathrm{BD}}} }
\newcommand{\DBDp}{ {\Delta_{\mathrm{BD}}^{\prime}} }
\newcommand{\tCBD}{ {\thicktilde{\mathcal{C}}_{\mathrm{BD}}} }
\newcommand{\tCBDp}{ {\thicktilde{\mathcal{C}}_{\mathrm{BD}}^{\prime}} }
\newcommand{\tDBD}{ {\thicktilde{\Delta}_{\mathrm{BD}}} }
\newcommand{\tDBDp}{ {\thicktilde{\Delta}_{\mathrm{BD}}^{\prime}} }
\newcommand{\RWo}{ {\mathrm{Re}[\mathcal{W}_0]} }
\newcommand{\thicktilde}[1]{\mathbf{\tilde{\text{$#1$}}}}

\def\smath#1{\text{\scalebox{.85}{$#1$}}}
\def\sfrac#1#2{\smath{\frac{#1}{#2}}}

\usepackage{footmisc} 

\usepackage{graphicx}
\usepackage{dcolumn}
\usepackage{bm}
\usepackage{color}
\usepackage{ulem}

\title{Hot Cosmic Qubits: Late-Time de Sitter Evolution and Critical Slowing Down}

\author[a,b]{Greg Kaplanek}
\author[a,b]{and C.P.~Burgess}
\affiliation[a]{Department of Physics \& Astronomy, McMaster University, Hamilton, Ontario, L8S 4M1, Canada}
\affiliation[b]{Perimeter Institute for Theoretical Physics, Waterloo, Ontario, N2L 2Y5, Canada }
\emailAdd{kaplaneg@mcmaster.ca}
\emailAdd{cburgess@perimeterinstitute.ca}

\date{}

\abstract {Temporal evolution of a comoving qubit coupled to a scalar field in de Sitter space is studied with an emphasis on reliable extraction of late-time behaviour. The phenomenon of critical slowing down is observed if the effective mass is chosen to be sufficiently close to zero, which narrows the window of parameter space in which the Markovian approximation is valid. The dynamics of the system in this case are solved in a more general setting by accounting for non-Markovian effects in the evolution of the qubit state. Self-interactions for the scalar field are also incorporated, and reveal a breakdown of late-time perturbative predictions due to the presence of secular growth.}

\begin{document}

\maketitle
\section{Introduction}
Many of the foundational paradoxes of gravitating quantum systems -- {\it e.g.}~black-hole information loss, eternal-inflation and multiverse issues -- arise due to puzzling behaviour displayed by simple systems at very late times. The simple systems used are often free quantum fields evolving in gravitational backgrounds, often spacetimes with horizons, that are chosen because explicit calculations can be made. When using these systems to make late-time inferences an implicit assumption is that it is the interaction with the background that always dominates, and any other neglected interactions can be treated as perturbations. 

Similarities between the physics of quantum systems in gravitational spacetimes (especially with horizons) and open systems \cite{StochInf, Starobinsky:1994bd, Tsamis:2005hd, OpenEFT1, OpenEFT2,OpenEFT3, OpenEFT4,OpenEFT5, OpenEFT6,OpenEFT7, OpenEFT8, Agon:2014uxa, Boyanovsky:2015tba, Boyanovsky:2015jen, Nelson:2016kjm, Hollowood:2017bil, Shandera:2017qkg, Agon:2017oia, Martin:2018zbe, Martin:2018lin} show that this assumption is actually unlikely to be true. The problem is that open systems (by definition) always have an `environment' whose properties are not measured (in this case, perhaps, the degrees of freedom behind the horizon). But a perturbative treatment of the interactions with such an environment essentially {\it always} fails at sufficiently late times. Ultimately it fails because the environment never goes away. Given enough time the effects of any interaction -- regardless of how weak it might be -- eventually accumulate to become large. Concretely, no matter how small an interaction Hamiltonian, $H_{\rm int}$, might be, there is eventually a time $t$ for which the evolution operator, $U(t) := \exp[-i (H_0 + H_{\rm int)} t]$, is not well-described by a finite number of powers of $H_{\rm int}$.

In practice this problem usually manifests itself through the phenomenon of `secular growth', where the coefficients in a perturbative expansion contain growing powers of time. For example if an observable $\cO(t)$ is computed in powers of a small coupling $g \ll 1$ 
\be
   \cO(t) = \sum_{n} c_n(t) \, g^n \,,
\ee
then the coefficients $c_n(t)$ typically grow without bound as $t$ gets large. Part of the motivation for using open-system tools \cite{OpenEFT1, OpenEFT2,OpenEFT3, OpenEFT4,OpenEFT5, OpenEFT6,OpenEFT7, OpenEFT8} is that they provide systematic ways to resum this late-time growth, often allowing perturbative results at short times to be converted into expressions that for large $t$ include all orders in $g^2 t$ but drop contributions of order $g^n t$ for $n > 2$. They thereby promise controlled and reliable approximations for late-time behaviour that straight-up perturbative methods cannot. 

This work accompanies a companion paper \cite{qubitpaper1}, which uses these tools to track the late-time evolution of an Unruh-DeWitt detector \cite{Unruh:1976db, DeWitt:1980hx}: a simple two-level system (or qubit) as it uniformly accelerates in flat space while coupled to a simple quantum scalar field (prepared in its Minkowski vacuum). Such a simple system allows these open-system tools to be explored in a very concrete and explicit way (see also \cite{Benatti:2004ee,Lin:2006jw,Yu:2008zza,Yu:2011eq,Hu:2011pd,Hu:2012ed,Fukuma:2013uxa,Menezes:2017rby,Tian:2016gzg,Moustos:2016lol,Menezes,Chatterjee:2019kxg}). Ref.~\cite{qubitpaper1} treats the field as an environment and integrates it out to set up the Nakajima-Zwanzig equation \cite{Nakajima,Zwanzig} describing its perturbative effects for the qubit. This is an integro-differential equation that is difficult to solve, but which simplifies at late times under certain assumptions to give an approximate late-time {\it Markovian} evolution. In particular much attention is given to the precise parameter range that controls this approximation. Not surprisingly the classic transition rates computed for Unruh-DeWitt detectors decades ago \cite{Unruh:1976db, DeWitt:1980hx, Sciama:1981hr} prove to break down at very late times. Evolution at much later times instead describes thermalization and decoherence as the qubit gets heated to the Unruh temperature. 

We here apply open-system tools to a similarly simple late-time question: what happens to such a qubit (again coupled to a scalar field) moving for very long times along a co-moving trajectory in de Sitter space. We again identify the relevant master equation for differential qubit evolution once the field is integrated out, and again find that a Markovian approximation works at sufficiently late times, asymptotically approaching a thermal state in much the same manner as in \cite{qubitpaper1}. de Sitter space brings an important complication, however: the length of time required for this Markovian limit to apply grows like an inverse power of the scalar mass if this mass is sufficiently small. That is, the qubit responds to field self-correlations that become increasingly persistent and extraordinarily long-lived. As a direct consequence of this, the qubit's approach to equilibrium becomes extremely slow in this regime (a phenomenon reminiscent of critical slowing down \cite{CritSlowDown1,CritSlowDown2}).

Another new feature of the de Sitter example is the emergence of a non-Markovian regime for which an approximate form of the late-time evolution can nevertheless be explicitly integrated to give a closed-form solution. This allows us to track a portion of the memory effects that work to foil the Markovian limit, and find a more general solution that settles to the expected late-time thermal state. We again develop precise conditions for when this solution is a good approximation, and recover the earlier Markovian solution as a limit. 

\section{Co-moving qubits and fields in de Sitter space}
 \label{sec:dSQubits}

This section reviews for later use some basic properties of de Sitter space, with details of the qubit/field system to be studied. The section then closes with a statement -- following \cite{qubitpaper1} -- of the Nakajima-Zwanzig equation that governs qubit evolution once the scalar field is integrated out.

The time evolution to be followed in later sections is along the coordinate direction within the flat slicing of de Sitter space, with line-element
\begin{eqnarray} \label{dSmetric}
 \exd s^2 \ = \ g_{\mu\nu} \exd x^{\mu} \exd x^{\nu } = \frac{ 1}{H^2 \eta^2} \Bigl( - \exd\eta^2 + \exd\mathbf{x}^2 \Bigr) =  - \exd \tau^2 + e^{2H\tau} \exd \bfx^2 \,,
\end{eqnarray}
where $H > 0$ is the Hubble constant and the conformal time $\eta$ is related to comoving time $\tau$ by $\eta = -H^{-1} e^{-H\tau}$ \cite{Weinberg}, with the negative sign chosen to ensure $\exd \eta / \exd \tau > 0$. The range $-\infty < \eta < 0$ corresponds to $- \infty < \tau < \infty$.

This metric is maximally symmetric and so has constant Ricci curvature\footnote{We use Weinberg's curvature conventions \cite{Weinberg}, that differ from those of Misner, Thorne and Wheeler \cite{MTW} only by an overall sign in the definition of the Riemann tensor.}
\begin{eqnarray}
\cR \ = \ - 12 H^2\,.
\end{eqnarray}

\subsection{Scalar fields in de Sitter space}

We consider a real scalar field with Lagrangian density
\begin{eqnarray}
\mathcal{L}(x) \ = \ - \frac{1}{2} \sqrt{-g} \bigg[ g^{\mu\nu} \partial_{\nu}\phi \; \partial_{\mu}\phi + m^2 \phi^2 + \xi \cR \phi^2 \bigg]
\end{eqnarray}
which includes a nonminimal interaction with the metric's Ricci scalar, $\cR$, with coupling parameter $\xi$. Because the Ricci scalar for de Sitter space is constant, from the point of view of the scalar field the nonminimal coupling effectively shifts the scalar field's mass from $m$ to
\begin{eqnarray}
M_{\mathrm{eff}}^2 = m^2 - 12 \xi H^2 \ .
\end{eqnarray}

The canonical Hamiltonian that generates the scalar field's evolution\footnote{This is a special instance of the Klein-Gordon Hamiltonian
\begin{eqnarray}
\cH & = & \int_{\Sigma_t} \exd^3\bx\ \sqrt{\gamma} \left[ \frac{1}{2} \dot{\phi}^2 + \frac{1}{2} \gamma^{ij}  \partial_i \phi\, \partial_j \phi + \frac{1}{2} m^2 \phi^2 \right] \notag 
\end{eqnarray}
for a static spacetime which admits a foliation with metric $\exd s^2 = - \exd t^2 + \gamma_{ij} \, \exd x^i \, \exd x^j$, with $\Sigma_{t}$ a sheet of fixed $x^0=t$.} in $\tau$ is therefore
\begin{eqnarray}
\cH & = & \int_{\Sigma_\tau} \exd^3\bx\ e^{3 H \tau} \left[ \frac{1}{2} \dot{\phi}^2 + \frac{1}{2} e^{- 2 H \tau} | \boldsymbol{\nabla} \phi |^2  + \frac{1}{2} M_{\mathrm{eff}}^2 \phi^2 \right] \label{freefield}
\end{eqnarray}
where $\Sigma_\tau$ is a sheet of fixed comoving time $\tau$. In what follows we take the scalar field to be prepared in the Bunch-Davies vacuum $\ket{{\mathrm{BD}}}$. 

Of particular interest for later evaluation of qubit evolution is the scalar field's Wightman function evaluated in this vacuum, which turns out to be given by \cite{Chernikov:1968zm,Schomblond:1976xc,Bunch:1978yq, CandelasRaine} 
\begin{eqnarray}
\braket{ \vacBD | \phi(\eta, \bx)\phi(\eta', \bx') | \vacBD } \ = \ \frac{H^2 (\tfrac{1}{4} - \nu^2)}{16 \pi\cos(\pi\nu)} \; {}_2F_1 \left( \tfrac{3}{2} + \nu, \tfrac{3}{2} - \nu ; 2 ; 1 + \frac{ (\eta - \eta' - i\epsilon )^2 - |\bx - \bx'|^2 }{4\eta\eta'}\right) \ \ \ \ \ \label{WBDgeneric}
\end{eqnarray}
where ${}_2F_1$ is the hypergeometric function and $\epsilon \to 0^{+}$ is an infinitesimal whose presence is determined by the Wightman boundary conditions and which determines how integrations should navigate around the singularity in the coincidence limit (where the points $(\eta, \mathbf{x})$ and $(\eta',\mathbf{x}')$ are lightlike separated). 

The parameter $\nu$ in \pref{WBDgeneric} is defined as the following function of $\xi$, $m$ and $H$:
\begin{eqnarray}
\nu := \sqrt{ \frac{9}{4} - \frac{M_{\mathrm{eff}}^2}{H^2} } \ = \ \sqrt{ \frac{9}{4} - \frac{m^2}{H^2} + 12 \xi } \ .
\end{eqnarray}
In our conventions the special case of a conformally coupled scalar field is the choice $m=0$ and $ \xi = - \tfrac{1}{6}$, in which case $M^2_{\mathrm{eff}} = 2 H^2$ and $\nu = \frac12$. A minimally-coupled massless (axion-like) field by contrast satisfies $m = \xi = M_{\rm eff} = 0$ and so $\nu = \frac32$. The Wightman function in the conformally coupled case is particularly simple, reducing to
\begin{eqnarray}
\braket{ \vacBD | \phi(\eta, \bx)\phi(\eta', \bx') | \vacBD } \ = \left(\frac{H^2 }{ 4 \pi^2 }\right) \frac{\eta \eta'}{ - ( \eta - \eta' - i \epsilon )^2 + |\bx - \bx'|^2}  \,.
\end{eqnarray}

\subsection{Qubit/field couplings}
\label{sec:setup}

To this field we couple a qubit, following the construction used in \cite{qubitpaper1}. The result is an Unruh-DeWitt detector coupled to a scalar field, along the lines of that first introduced in \cite{Unruh:1976db,DeWitt:1980hx}. We take the qubit's free Hamiltonian to be
\begin{eqnarray}
\mathfrak{h}_0 \ = \ \frac{\omega_0}{2} \, \boldsymbol{\sigma_{3}} \ = \ \left[ \begin{matrix} {\omega_0}/{2} & 0 \\ 0 & - {\omega_0}/{2} \end{matrix} \right] \,,
\end{eqnarray}
where $\omega_0 > 0$ denotes the splitting between the two qubit energies. We suppose the qubit sits indefinitely at a fixed position $\mathbf{y} \in \bR^{3}$, and so follows the (geodesic) trajectory of a co-moving observer
\begin{eqnarray}
y^\mu(\tau) \ = \ [ \eta(\tau), \mathbf{y} ] \ = \ [ - H^{-1} e^{- H \tau}, \mathbf{y} ] \label{trajectory}
\end{eqnarray}
along which the coordinate $\tau$ is the proper time as measured with the spacetime metric of \pref{dSmetric}. 

The Hilbert space of states for the combined qubit/field system is the product of the Fock space for the field with the qubit's $2\times 2$ space of states. The free Hamiltonian acting on this Hilbert space is then
\begin{eqnarray}
H_{0} & = & \cH \otimes \mathbf{I} + \mathcal{I} \otimes \mathfrak{h}_0 \,,
\end{eqnarray}
where $\boldsymbol{I}$ and $\mathcal{I}$ are identity operators.\footnote{Because $H_0$ generates translations in comoving time $\tau$, and there is no need for a time-dilation factor as in \cite{qubitpaper1}.} The total Hamiltonian is $H = H_{0} + H_{{\rm int}\,0}$ where the qubit/field coupling is described by the interaction Hamiltonian 
\begin{eqnarray}
  H_{\mathrm{int}\,0} & = & g \, \phi\big[ y(\tau)\big] \otimes \mathfrak{m}  \ , \label{Hint0}
\end{eqnarray}
and the dimensionless coupling $0 < g \ll 1$ is small enough to justify a perturbative treatment. We follow common convention and choose $\mathfrak{m} = \boldsymbol{\sigma_{1}}$, but all that really counts is that $\mathfrak{m}$ and $\mathfrak{h}$ do not commute with one another so that $H_{{\rm int}\,0}$ drives transitions between the zeroeth-order qubit energy eigenstates. 

Naively one would perform perturbation theory simply by expanding in powers of $H_{{\rm int}\,0}$. It happens, however, that at $\mathcal{O}(g^2)$ the qubit/field interaction shifts the qubit energy splitting so that $\omega := E_\uparrow - E_\downarrow = \omega_0 + g^2 \omega_1$ for calculable $\omega_1$. A better choice for perturbation theory uses the physical value $\omega$ in the unperturbed Hamiltonian, and so writes
\be 
H_{\rm free}  =  \cH \otimes \mathbf{I} + \mathcal{I} \otimes \mathfrak{h} \quad \hbox{with} \quad
\mathfrak{h} \ = \ \frac{\omega}{2} \, \boldsymbol{\sigma_{3}}   \,.
\ee
With this choice $H = H_{0} + H_{{\rm int}\,0} = H_{\rm free} + H_{\rm int}$ and so the interaction Hamiltonian acquires a new counter-term, with
\begin{eqnarray}
  H_{\mathrm{int}} & = & g \, \phi\big[ y(\tau)\big] \otimes \mathfrak{m}  \ + \ \mathcal{I} \otimes \frac{g^2 \omega_1}{2}  \boldsymbol{\sigma_{3}} \,. \label{Hint}
\end{eqnarray}
Although not motivated by divergences, this counter-term interaction has an added benefit inasmuch as the quantity $\omega_1$ happens to cancel an ultraviolet divergence that arises at second order in $g$.

\subsection{Time evolution and the Nakajima-Zwanzig equation}
\label{sec:NZeq}

In principle the question of calculating the system's time evolution perturbatively in powers of the coupling $g$ is a solved problem. One converts to the interaction picture, by performing a unitary transformation, $O \to O^\ssI := U^\dagger(\tau)\, O \,U(\tau)$, for any operator $O$, with
\begin{eqnarray}
U_0(\tau) \ = \ \mathcal{T}\exp\left( - i \int_0^\tau \exd s\ H_0 \right) \ = \ e^{- i \cH \tau} \otimes e^{- i \mathfrak{h} \tau} \,.
\end{eqnarray}
As applied to the system's state (described by its density matrix, $\rho$) this transformation removes the `free' part of the evolution, leaving $\rho$ to be evolved by the interaction-picture Liouville equation
\be
 \frac{\partial \rho^{\ssI}}{\partial \tau} = -i \Bigl[ V(\tau) \,, \rho^\ssI(\tau) \Bigr] \,, \label{Liouville}
\ee
where
\begin{eqnarray}
V(\tau) \ = \ U^{\dagger}_0(\tau) H_{\mathrm{int}} U_0(\tau) \ = \ g \, \phi^\ssI[y (\tau)] \otimes \mathfrak{m}^\ssI(\tau)  \,.
\end{eqnarray}
Here $\phi_\ssI (\tau, \bfx) := e^{+ i \cH \tau} \phi(\bfx) e^{- i \cH \tau}$ is the interaction-picture field and the interaction-picture qubit coupling matrix is 
\be
  \mathfrak{m}^\ssI(\tau) := e^{+ i \mathfrak{h} \tau} \mathfrak{m} \, e^{- i \mathfrak{h} \tau} \,. 
\ee

Standard arguments then solve eq.~\pref{Liouville} interatively to any required power of $V(\tau)$, subject to an initial condition, $\rho(0)$, at $\tau = 0$. In what follows we take the field to be initially prepared in the Bunch-Davies vacuum $\ket{ \mathrm{BD} }$, and assume the initial qubit state to be uncorrelated with the field degrees of freedom:
\begin{eqnarray}
\rho(0) \ = \ \ket{ \mathrm{BD} } \bra{ \mathrm{BD} } \otimes \boldsymbol{\varrho_0} \label{uncorrelated1}
\end{eqnarray}
where $\boldsymbol{\varrho_0}$ is the qubit's initial $2\times 2 $ hermitian density matrix, satisfying $\tr \boldsymbol{\varrho_0} = 1$. 

But a complete solution for $\rho(\tau)$ is overkill if the goal is simply to predict the behaviour of qubit observables, with no measurements made in the scalar-field sector. For such observables the time-evolution problem is completely solved if the time-dependence of the reduced density matrix,
\begin{eqnarray} \label{RedRhoDef}
\boldsymbol{\varrho}(\tau) := \underset{\phi}{\mathrm{Tr}}\big[ \rho(\tau)\big]  \,,
\end{eqnarray}
is known, with initial condition
\be \label{varrhoIC}
  \boldsymbol{\varrho}(0) = \boldsymbol{\varrho_0} \,.
\ee
In \pref{RedRhoDef} the partial trace is only over the field-theory sector (and not also the qubit). 

The Nakajima-Zwanzig equation provides a formal solution to the problem of identifying the evolution equation for $\boldsymbol{\varrho}(\tau)$ that should replace \pref{Liouville}. It is obtained by perturbatively solving for the evolution of the scalar-field state and using this to eliminate the scalar completely\footnote{See Appendix A of \cite{qubitpaper1} for more details of how this is done.} from \pref{Liouville}, leaving an evolution equation that involves only $\boldsymbol{\varrho}$. 

When the dust settles, a calculation identical to that in  \cite{qubitpaper1} shows the qubit's interaction-picture reduced density matrix 
\begin{eqnarray}
\boldsymbol{\varrho}^\ssI (\tau)  :=   e^{+ i \mathfrak{h} \tau } \boldsymbol{\varrho}(\tau) \, e^{- i \mathfrak{h} \tau} \label{INTpicturereducedstate}
\end{eqnarray}
evolves -- at $\mathcal{O}(g^2)$ -- according to the following Nakajima-Zwanzig equation:
\begin{eqnarray}
\frac{\partial \boldsymbol{\varrho}^\ssI (\tau)}{\partial \tau} & \simeq & g^2 \int_0^\tau \exd s\ \bigg( \WBD(\tau - s) \big[ \mathfrak{m}_\ssI(s) \, \boldsymbol{\varrho}^\ssI (s), \mathfrak{m}_\ssI(\tau) \big] + \WBD(\tau - s)^{\ast} \big[ \mathfrak{m}_\ssI(\tau) , \boldsymbol{\varrho}^\ssI (s)\, \mathfrak{m}_\ssI(s)  \big] \bigg) \label{INTpictureNZ}  \quad \quad \\
& \ & \quad \quad \quad \quad \quad \quad \quad \quad \quad \quad \quad \quad \quad \quad \quad \quad \quad \quad \quad \quad \quad \quad \quad - i \left[ \frac{g^2 \omega_1}{2} \boldsymbol{\sigma_{3}} , \boldsymbol{\varrho}^{\ssI}(\tau) \right] \, . \notag 
\end{eqnarray}
Here the function $\WBD$ denotes the Bunch-Davies Wightman function $\braket{ \mathrm{BD} | \phi(x) \phi(x') | \mathrm{BD} }$, evaluated along the qubit's trajectory
\begin{eqnarray} 
  \WBD(\tau_1 - \tau_ 2) & := & \braket{ \vacBD | \phi(\tau_1, \bfy) \phi( \tau_2, \bfy) |\vacBD } \, . \label{WightmanDef}
\end{eqnarray}
It is the symmetry of the spacetime and the choice of the trajectory \pref{trajectory} that ensures that the Wightman function depends only on the difference $\tau_2 - \tau_1$.

In components, after some matrix algebra \pref{INTpictureNZ} yields the equations of motion
\begin{eqnarray}
\frac{\partial \varrho^{\ssI}_{11}}{\partial \tau} & = & g^2 \int_{-\tau}^{\tau} \exd s \; \WBD(s) \, e^{- i \omega s} - 4 g^2 \int_0^{\tau} \exd s\ \mathrm{Re}[\WBD(s) ] \, \cos(\omega s) \varrho^{\ssI}_{11}(\tau - s)  \, , \label{rho111} \\
\frac{\partial \varrho^{\ssI}_{12}}{\partial \tau} & = & - i g^2 \omega_1 \; \varrho^{\ssI}_{12}(\tau) - 2 g^2 \int_0^\tau \exd s \ \mathrm{Re}[\WBD(s)] e^{+ i \omega s} \varrho_{12}^{\ssI}(\tau - s) \ \label{rho121} \\
&\ & \quad \quad  \quad \quad  \quad \quad  \quad \quad \quad  \quad \quad  \quad \quad + \ 2 g^2  e^{+ 2 i \omega \tau} \int_0^\tau \exd s \ \mathrm{Re}[\WBD(s)] e^{- i \omega s} \varrho_{12}^{\ssI\ast}(\tau - s)\ . \notag 
\end{eqnarray}
where the integration variable are also switched, $s \to \tau - s$. The properties $\tr\boldsymbol{\varrho}(\tau) = 1$ and $\boldsymbol{\varrho}^{\dagger}(\tau) = \boldsymbol{\varrho}(\tau)$ have also been used to eliminate $\varrho_{22}(\tau) = 1 - \varrho_{11}(\tau)$ and $\varrho_{21}(\tau) = \varrho_{12}^{\ast}(\tau)$. Eqs.~\pref{rho111} and \pref{rho121} make it clear that the diagonal and off-diagonal components of $\boldsymbol{\varrho}$ evolve independent of one other.

\subsection{The Wightman function along the qubit trajectory}
\label{sec:Wightman}

The Nakajima-Zwanzig equation boils all effects of the scalar field to the correlations encoded in the Wightman function $\WBD(\tau)$ evaluated along the qubit trajectory \pref{trajectory}. Specializing \pref{WBDgeneric} to this trajectory gives the explicit form
\begin{eqnarray}
\mathcal{W}_{\mathrm{BD}}(\tau) \ = \ \frac{H^2 (\tfrac{1}{4} - \nu^2)}{16 \pi\cos(\pi\nu)} \; _2F_1 \left( \tfrac{3}{2} + \nu, \tfrac{3}{2} - \nu ; 2 ; 1 + \big[ \sinh( \tfrac{H\tau}{2} ) - i \tfrac{H \epsilon}{2} \big]^2 \right) \,.\ \ \ \ \ \label{WightmanBD}
\end{eqnarray}
where we remind the reader that $\epsilon \to 0^{+}$ is taken at the end of any calculation. As is easy to verify this function satisfies the identities
\begin{eqnarray}
\WBD^{\ast} (\tau) \ = \ \WBD(-\tau) \label{WBDsymmetry}
\end{eqnarray}
and
\begin{eqnarray}
\WBD\big(\tau - {2 \pi i } / {H}\big) \ = \ \WBD(-\tau) \,. \label{KMS}
\end{eqnarray}
This last condition -- the Kubo-Martin-Schwinger (KMS) \cite{Kubo:1957mj,Martin:1959jp} condition -- expresses detailed balance and ultimately ensures that $\boldsymbol{\varrho}(\tau)$ eventually asymptotes to a thermal state with Gibbons-Hawking temperature\footnote{We remark that if extending the domain of $\WBD(\tau)$ within an integral to include $\tau< 0$ then one must choose $\WBD(\tau) \simeq ( \RWo + i \mathrm{sgn}(\tau) \mathrm{Im}[\mathcal{W}_0] ) e^{- \kappa |\tau|}$ so as to preserve the important property \pref{WBDsymmetry}. }  \cite{Gibbons:1977mu}
\be
   T = \frac{H}{2\pi} \label{GibbonsHawking}\,.
\ee

In the special case of a conformal scalar case -- {\it i.e.}~where $M_{\mathrm{eff}}^2 = 2H^2$ -- the Wightman function simplifies to
\begin{eqnarray}
\mathcal{W}_{\mathrm{BD}}(\tau) \ = \  - \frac{H^2}{16 \pi^2} \frac{1}{\big[ \sinh(\tfrac{H\tau}{2}) - i \frac{H\epsilon}{2} \big]^2}\qquad \hbox{(conformal scalar)}  \,, \label{conformalWBD}
\end{eqnarray}
which (after replacing $H$ with the acceleration parameter) agrees with the Wightman function for a massless field in Minkowski space evaluated along a uniformly accelerated trajectory \cite{Takagi:1986kn,Birrell:1982ix, qubitpaper1}.

As discussed in \cite{qubitpaper1}, for the present purposes (late-time evolution) what is important about expressions \pref{WightmanBD} and \pref{conformalWBD} is their asymptotic form in the limit $H \tau \gg 1$. As is shown (for various choices of parameters) in Table \ref{BDasymptotics} the function $\mathrm{Re}[\WBD(\tau)]$ always falls off exponentially, $\WBD \propto e^{- \tau / \tau_c}$, for large enough $\tau$. This falloff is important because it makes possible the existence of a simpler Markovian limit, at least for times, $\tau \gg \tau_c$. 

\begin{table}[h] 
  \centering    
     \centerline{\begin{tabular}{ r|l|l|c| }
 \multicolumn{1}{r}{}
 & \multicolumn{1}{c}{$\nu$}
 & \multicolumn{1}{c}{$M_{\mathrm{eff}}/H$}
 & \multicolumn{1}{c}{$\mathrm{Re}[\mathcal{W}_{\mathrm{BD}}(\tau)]$} \\
\cline{2-4}
$\ $ & $\nu = i \mu$, $\mu > 0$ & $ \frac{ M_{\mathrm{eff}}}{H} \in \left( \frac{3}{2}, \infty \right) $ & $\stackrel{\ }{\underset{\ }{ \dfrac{H^2\sqrt{1+4\mu^2} \sqrt{\tanh(\pi\mu)}}{4\pi^{3/2}\sqrt{\mu}} \; e^{-\tfrac{3}{2} H \tau } \sin\big( \mu H \tau + \mathrm{Arg}\left[ \Gamma(\frac{3}{2} - i \mu) \Gamma(i\mu) \right] \big) } }$  \\
\cline{2-4}
$\ $ &
$\nu \in \left[0, \frac{3}{2} \right) \setminus \{1\}$ & $\frac{ M_{\mathrm{eff}}}{H} \in \left( 0, \frac{3}{2} \right] \setminus \left\{ \frac{\sqrt{5}}{2} \right\} \ $ & $\stackrel{\ }{\underset{\ }{ - \dfrac{H^2}{4\pi^{5/2}} \sin(\pi\nu) \Gamma(\tfrac{3}{2} - \nu) \Gamma(\nu ) e^{- \left( \tfrac{3}{2} - \nu \right) H \tau} } }$ \\
\cline{2-4}
$\ $ &
$\nu = 1 $  & $\frac{ M_{\mathrm{eff}}}{H} = \frac{\sqrt{5}}{2}  $ & $\stackrel{\ }{\underset{\ }{- \dfrac{3H^2}{8\pi} e^{-\tfrac{5}{2}H\tau} }}$  \\
\cline{2-4}
    \end{tabular} }
        \caption{Leading asymptotic behaviour of $\mathrm{Re}[\mathcal{W}_{\mathrm{BD}}(\tau)]$ in the limit $H \tau \gg 1 $ for several possible values of $M_{\eff}/H$ (details given in Appendix \ref{App:WBDLargeTime}). In the first row $\nu = i \mu$ lies on the positive imaginary axis with $\mu \equiv \sqrt{ M_{\eff}^2/H^2 - 9/4 } >0$ ensured because $M_{\mathrm{eff}}/H > \frac{3}{2} $. The leading coefficient in the asymptotic expansion for $\nu \in [0, 1 ) \cup (1,\frac{3}{2})$ vanishes when $\nu \to 1^{\pm}$ which is why there is a separate row for the special case of $\nu=1$.} \label{BDasymptotics}
\end{table}

We see from this Table that -- with one exception -- the time-scale $\tau_c$ over which $\mathrm{Re}[\mathcal{W}_{\mathrm{BD}}(\tau)]$ is exponentially suppressed is generically of order the Hubble time. The exception is close to the massless, minimally coupled case, for which $M_{\rm eff} \to 0$ and $\nu \to \frac32$. Writing the asymptotic form as
\begin{eqnarray}
\WBD(\tau) & \simeq & \mathcal{W}_0 \; e^{ - \kappa \tau} \quad \quad \quad \quad (\mathrm{when\ } 0 < M_{\mathrm{eff}} < \tfrac{3}{2} H \ \mathrm{and}\  H\tau \gg 1) \,, \label{WBDsmallMeff}
\end{eqnarray}
provided $M_{\rm eff}^2 \neq \frac52 \,H^2$ the parameters $\cW_0$ and $\kappa$ are given by (see Appendix \ref{App:WBDLargeTime} for details)
\begin{eqnarray}
\mathcal{W}_0 \ : = \ \sfrac{H^2}{4\pi^{5/2}} i e^{ i \pi \nu} \Gamma(\tfrac{3}{2} - \nu) \Gamma(\nu) \quad
\hbox{and} \quad \kappa \  :=  \ \left( \tfrac{3}{2} - \nu \right) H  \,. \label{W0def}
\end{eqnarray}
For the limit $M_{\rm eff} \to 0$ both $\cW_0$ and $\kappa^{-1}$ blow up, with asymptotic forms
\be 
\mathrm{Re}[\WBD(\tau)]  \simeq  \left( \frac{3H^4}{8\pi^2 M_{\mathrm{eff}}^2} \right) \; \exp\left[ - \sfrac{M_{\mathrm{eff}}^2 \tau}{3H}  \right] \quad \quad \quad \quad (\mathrm{when\ } 0 < M_{\mathrm{eff}} \ll H \ \mathrm{and}\  H\tau \gg 1) \ .
\ee
Evidently in the regime $0 < M^2_{\mathrm{eff}} \lesssim 2H^2$ the function $\mathrm{Re}[\mathcal{W}_{\mathrm{BD}}(\tau)]$ falls off much more slowly, with the longer-range correlations noted in \cite{Linde:1984ir,Fukuma:2013uxa,Starobinsky:1994bd}.

\begin{figure}[h]
\begin{center}
\includegraphics[scale=0.30]{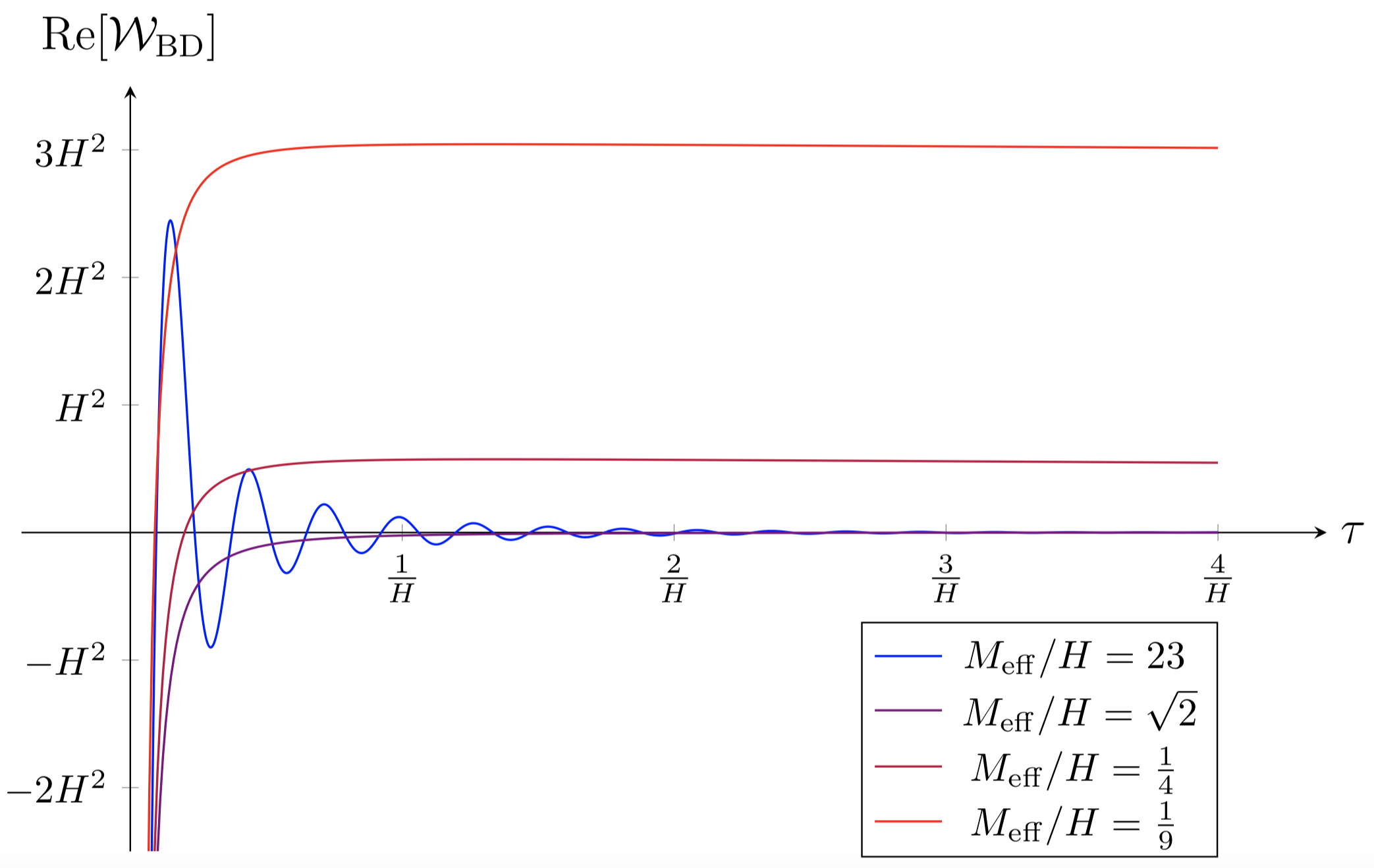}
\caption{A plot of $\mathrm{Re}[\mathcal{W}_{\mathrm{BD}}]$ versus $\tau$ for various values of $M_{\mathrm{eff}}/H$. The oscillatory component visible for $M_{\rm eff}/H > \frac32$ disappears once $M_{\rm eff} /H < \frac32$. The exponential damping occurs over a few Hubble times ($H^{-1}$) for $M^2_{\mathrm{eff}}   \gtrsim 2 H^2$ but becomes much longer as $M_{\rm eff} \to 0$, becoming of order $H/M^2_{\rm eff}$ in the regime $M^2_{\mathrm{eff}} \ll H^2$.} \label{fig:WBD} 
\end{center}
\end{figure}

For later sections it is also useful to know the subdominant terms in the expansions about the $H \tau \to \infty$ and $M_{\rm eff}/H \to 0$ limits. The next-to-leading terms in the $M_{\rm eff}/H$ expansion are
\begin{eqnarray}
\kappa & \simeq & \frac{M_{\mathrm{eff}}^2}{3H} + \frac{M_{\mathrm{eff}}^4}{27H^3} + \ldots\qquad\qquad\qquad \hbox{(if $M_{\rm eff} \ll H$)} \ , \label{kappaExpansion} \\
\mathrm{Re}[\mathcal{W}_0] & \simeq & \frac{3H^4}{8\pi^2 M_{\mathrm{eff}}^2} - \frac{(7-6\ln 2)H^2}{24\pi^2} + \ldots \quad \hbox{(if $M_{\rm eff} \ll H$)} \ \ .\label{RWoExpansion}
\end{eqnarray}
For $H\tau \gg 1$, if the sub-leading corrections to the late-time asymptotic series for $\WBD(\tau)$ are defined by
\begin{eqnarray}
f(\tau) \ := \  \WBD(\tau) -  \mathcal{W}_0 \; e^{ - \kappa \tau}  \ , \label{subleading}
\end{eqnarray}
then $f(\tau)$ falls much faster with $\tau$ than does $\WBD(\tau)$ in the limit of vanishing effective mass. Sub-leading corrections to $f(\tau)$ in \pref{subleading} are $\mathcal{O}(e^{- H \tau})$ even in the $M_{\mathrm{eff}} / H \ll 1$ limit.\footnote{The sub-leading corrections are $\mathcal{O}(e^{- \left({5}/{2} - \nu \right) H \tau}) \simeq \mathcal{O}(e^{- H\tau})$ when $M_{\mathrm{eff}} / H \ll 1$. See \pref{WBDleadingsmall} in Appendix \ref{App:WBDLargeTime}.}  These asymptotic properties are also visible in the numerical plots given in Fig.~\ref{fig:WBD}.

\section{Late-time Markovian behaviour}
\label{sec:MarkovianSection}

This section exploits the exponential falloff of the Wightman function, $\WBD(\tau) \simeq \mathcal{W}_0 \; e^{ - \kappa \tau} $, to derive an approximate form for eqs.~\pref{rho111} and \pref{rho121} that captures well the late-time evolution. 

\subsection{Markovian approximation}
\label{sec:MarkovApprox}

The idea behind the approximation is simple: to the extent that one's interest is in slow evolution at very late times ({\it i.e.} $\tau \gg \tau_c$) then $\boldsymbol{\varrho}(\tau)$ does not vary significantly over the comparatively short time-scales over which the integrals in \pref{rho111} and \pref{rho121} have appreciable support. It should be a good approximation in this regime to Taylor expand 
\be
 \boldsymbol{\varrho}(\tau - s) \simeq \boldsymbol{\varrho}(\tau) - s \, \partial_\tau \boldsymbol{\varrho}(\tau) + \cdots \,,
\ee
within the integrand \cite{Montroll,Barnett}, in which case the dependence on $\boldsymbol{\varrho}(\tau)$ comes out of the integral. Keeping only the leading term in this expansion allows \pref{rho111} and \pref{rho121} to be rewritten in the form
\begin{eqnarray}
\frac{\partial \varrho^{\ssI}_{11}}{\partial \tau} & \simeq & g^2 \int_{-\tau}^{\tau} \exd s \; \WBD(s) \, e^{- i \omega s} - 4 g^2 \int_0^{\tau} \exd s\ \mathrm{Re}[\WBD(s) ] \, \cos(\omega s) \varrho^{\ssI}_{11}(\tau)  \, , \label{rho11two} \\
\frac{\partial \varrho^{\ssI}_{12}}{\partial \tau} & \simeq & - i g^2 \omega_1 \; \varrho^{\ssI}_{12}(\tau) - 2 g^2 \int_0^\tau \exd s \ \mathrm{Re}[\WBD(s)] e^{+ i \omega s} \varrho_{12}^{\ssI}(\tau) \ \label{rho12two} \\
&\ & \quad \quad  \quad \quad  \quad \quad  \quad \quad \quad  \quad \quad  \quad \quad + \ 2 g^2  e^{+ 2 i \omega \tau} \int_0^\tau \exd s \ \mathrm{Re}[\WBD(s)] e^{- i \omega s} \varrho_{12}^{\ssI\ast}(\tau)\ . \notag 
\end{eqnarray}
As quantified below, the error in dropping subdominant terms in the integrand should be of order $(\tau_c \, \partial_\tau)^n \boldsymbol{\varrho}(\tau)$, and therefore should be small when $\boldsymbol{\varrho}$ only varies slowly over times of order $\tau_c$. 

Specialization to times $\tau \gg \tau_c$ allows further simplification because in this regime the limits of integration in \pref{rho11two} and \pref{rho12two} can also be taken to infinity up to exponentially small corrections. Switching back to the Schr\"odinger picture, this allows the above equations to be written as
\begin{eqnarray}
\frac{\partial \varrho_{11}}{\partial \tau} & \simeq & g^2 \RBD - 2 g^2 \CBD \varrho_{11}(\tau)  \, , \label{rho11three} \\
\frac{\partial \varrho_{12}}{\partial \tau} & \simeq & - i \omega \varrho_{12}(\tau) - g^2 \CBD \varrho_{12}(\tau) + g^2 (\CBD - i \DBD) \varrho_{12}^{\ast}(\tau) \,.\label{rho12three}
\end{eqnarray}
Here the coefficient functions are defined by the integrals
\begin{eqnarray}
\CBD(\omega) &:=& 2 \int_0^\infty \exd s\ \mathrm{Re}[\WBD(s)] \cos(\omega s) \label{CBDdef} \\
  & = & \frac{H}{4\pi^3} \cosh\left( \frac{\pi \omega}{H} \right) \; \left|  \Gamma\left( \frac{3}{4} + \frac{\nu}{2} + i \frac{\omega}{2H} \right) \Gamma\left( \frac{3}{4} - \frac{\nu}{2} + i \frac{\omega}{2H} \right) \right|^2   \,,\nn
\end{eqnarray}
and \cite{Higuchi:1986ww,Garbrecht:2004du}
\begin{eqnarray}
\RBD(\omega) & : = & \int_{-\infty}^{\infty} \exd s \ \WBD(s) e^{- i \omega s}    \label{RBDdef} \\
& = & \frac{H}{4\pi^3} \; e^{- \tfrac{\pi\omega}{H}} \left|  \Gamma\left( \frac{3}{4} + \frac{\nu}{2} + i \frac{\omega}{2H} \right) \Gamma\left( \frac{3}{4} - \frac{\nu}{2} + i \frac{\omega}{2H} \right) \right|^2\,,\nn
\end{eqnarray}
while 
\begin{eqnarray}
\DBD(\omega) &:=& 2 \int_0^\infty \exd s\ \mathrm{Re}[\WBD(s)] \sin(\omega s) \label{DBDdef} \,.
\end{eqnarray}
The function $\DBD$ diverges at the $s \to 0$ end of the integration, and its form is evaluated more explicitly in Appendix \ref{App:DBD}. We choose to regulate this divergence by making the small-distance regulator $\epsilon$ in the Wightman function \pref{WightmanBD} small but finite, leading to a divergence that is logarithmic in $\epsilon$ whose form is derived more explicitly in Appendix \ref{App:DBD}.

Notice also that the disappearance of the quantity $\omega_1$ when passing from \pref{rho12two} to \pref{rho12three} occurs because we choose $\omega_1 = - \DBD$ to cancel a term $i \DBD \, \varrho_{12}$ in \pref{rho12three} in order to ensure that $\omega$ appears in the same way in the evolution of $\boldsymbol{\varrho}$ as should the physical qubit energy splitting. 

The identities \pref{WBDsymmetry} and \pref{KMS} impose some relations amongst the above integrals. First, \pref{WBDsymmetry} implies that $\mathrm{Re}[\WBD(\tau)]$ and $\mathrm{Im}[\WBD(\tau)]$ are even and odd functions in $\tau$, respectively, and so 
\be 
\RBD(\omega)  =  \CBD(\omega) + \SBD(\omega) \,,
\ee
where
\be
\SBD(\omega) := 2 \int_0^\infty \exd s\ \mathrm{Im}[\WBD(s)] \sin(\omega s) \,.\label{SBDdef}
\ee
Furthermore, \pref{KMS} implies $\RBD$ satisfies  
\begin{eqnarray}
\RBD(\omega) - e^{ - 2 \pi  \omega / H } \RBD(- \omega) \ = \ 0 \ , 
\end{eqnarray}
which is the detailed-balance relation \cite{Takagi:1986kn} (and so underlies the thermal nature of many of the late-time equilibrium properties). Because $\CBD$ and $\SBD$ are even and odd in $\omega$ respectively, the detailed-balance relation also implies
\begin{eqnarray}
\frac{\SBD(\omega)}{\CBD(\omega)} \ = \ - \tanh\left( \frac{\pi \omega}{H} \right) \ , \label{ratioSC}
\end{eqnarray}
from which we see
\begin{eqnarray}
\RBD(\omega) \ = \ \frac{2}{e^{2 \pi \omega / H}+1} \; \CBD(\omega) \ = \ \frac{2}{e^{2 \pi \omega / H} - 1} \;\SBD(\omega) \ . \label{RrelatedCS}
\end{eqnarray}

\subsection{Late-time evolution}
\label{ssec:LateTimeEv}

The quantity $\RBD$ appears throughout the literature on Unruh-DeWitt detectors \cite{Takagi:1986kn,Sciama:1981hr,Birrell:1982ix,DeWitt:1980hx} because it governs the perturbative excitation rate of a qubit that is initially prepared in its ground state: $\boldsymbol{\varrho_0} = \boldsymbol{\varrho}_{\rm vac}$, where
\be
   \boldsymbol{\varrho}_{\rm vac} := \ket{\downarrow}\bra{\downarrow} = (\boldsymbol{I} - \boldsymbol{\sigma_{3}})/2 \,.
\ee
With this choice $\varrho_{11}(0) = \varrho_{12}(0) = 0$ and so both $\varrho_{11}(\tau)$ and $\varrho_{12}(\tau)$ are at most of order $g$ at later times. Consequently any appearance of these quantities on the right-hand-side of eqs.~\pref{rho11three} and \pref{rho12three} can be dropped if we wish to compute $\partial_\tau \boldsymbol{\varrho}(\tau)$ only to order $g^2$. This leads to the 
standard lowest-order prediction for the rate of accumulation of probability into the excited qubit state:
\begin{eqnarray}
  \frac{\partial {\varrho}_{11}(\tau)}{\partial \tau} \simeq g^2 \RBD \ . \quad \quad \quad \label{ratePT}
\end{eqnarray}

As is clear from the above derivation, this rate strictly only applies at times $\tau \gg \tau_c$ and in this sense is usually interpreted as the qubit's late-time transition rate. However it is also clear that the prediction \pref{ratePT} cannot apply at asymptotically late times. Eq.~\pref{ratePT} must break down at sufficiently late times because unitarity requires $0 \leq \varrho_{11} (\tau) \leq 1$ for all $\tau$ and so forbids eternally accumulating probability into the state $\ket{\uparrow}$ with constant rate.\footnote{In this qubit evolution differs from standard discussions of exponential decays of unstable states. Decays can proceed indefinitely with constant rate because decay daughters can escape to infinity and so do not accumulate probability in the same way as does a qubit.} 

What happens at these later times? This is where eqs.~\pref{rho11three} and \pref{rho12three} prove their worth, because their domain of validity can extend to much later times than their derivation naively would indicate \cite{OpenEFT2, qubitpaper1}. Their extended domain of validity arises because -- unlike for (say) eqs.~\pref{rho111} and \pref{rho121} or eqs.~\pref{rho11two} and \pref{rho12two} -- neither of \pref{rho11three} or \pref{rho12three} make explicit reference to the initial time at which the evolution starts. This means that although they were derived starting from an assumed initial state at $\tau = 0$, they could have equally well been derived starting at some other initial time, $\tau_1$ or $\tau_2$. Although the derivation that starts from $\tau= \tau_n$ is only valid over a limited window of time, $\tau \in \cS_n$, to the future of $\tau_n$, since it is the same differential equation that is derived for all $n$, the differential equations themselves -- {\it i.e.}~eqs.~\pref{rho11three} and \pref{rho12three} -- remain valid for the {\it union} of all of these intervals: $\cS = \cup_n \cS_n$. 

This argues that at much later times it is eqs.~\pref{rho11three} and \pref{rho12three} that govern the evolution of $\boldsymbol{\varrho}(\tau)$, and it is the presence of the other terms on their right-hand-side beyond the $g^2 \RBD$ term that build in the constraints of unitarity that are missing in \pref{ratePT}. Furthermore, keeping only terms to $\cO(g^2)$ in \pref{rho11three} and \pref{rho12three} implies that their solutions can be trusted to all orders in $g^2 \tau$ as $g \to 0$ and $\tau \to \infty$, though they do not properly capture terms of order $g^n \tau$ with $n > 2$. This is most easily seen by scaling $\tau \to \hat \tau := g^2 \tau$ in eqs.~\pref{rho11three} and \pref{rho12three}. We return to a more careful determination of the domain of validity of eqs.~\pref{rho11three} and \pref{rho12three} below. 

What do these equations imply for late-time evolution? Eq.~\pref{rho11three} has the solution
\be 
\varrho_{11}(\tau)  =  \frac{1}{e^{ 2\pi \omega / H } + 1} + \left[ \varrho_{11}(0) - \frac{1}{e^{ 2\pi \omega / H } + 1} \right] e^{ - 2 g^2 \CBD \tau } \label{Solution11Schro}
\ee
where the relation \pref{RrelatedCS} has been used to eliminate $\RBD$. Eq.~\pref{rho12three} similarly integrates to give
\be 
\varrho_{12}(\tau)  =  e^{-g^2 \CBD \tau} \left\{ \varrho_{12}(0) \left[ \cos(\Sigma\tau) - i \frac{\omega}{\Sigma} \sin(\Sigma \tau) \right] + \varrho_{12}^{\ast}(0)\, \frac{g^2 \CBD - i g^2 \DBD}{\Sigma} \;\sin(\Sigma\tau) \right\} \label{exact12Schro}
\ee
where
\be
\Sigma  =  \sqrt{ \omega^2  - g^4(\mathcal{C}_{\mathrm{BD}}^2 + \Delta_{\mathrm{BD}}^2)  } \ .
\ee

As later sections show, the domain of validity of \pref{exact12Schro} is slightly more complicated to justify in detail because of the potential disparity in scale between the qubit gap $\omega$ and the $\mathcal{O}(g^2)$ factors. The discussion is simplest in the `non-degenerate' regime, for which
\begin{eqnarray}
\omega \gg g^2 \sqrt{ \mathcal{C}_{\mathrm{BD}}^2 + \Delta_{\mathrm{BD}}^2 } \ , \label{g2constraint}
\end{eqnarray}
however, because then $\cO(\omega)$ and $\cO(g^2)$ effects do not interfere with one another. We therefore assume \pref{g2constraint} to be true in what follows (though the limit of smaller $\omega$ goes through similarly, as described in \cite{qubitpaper1}). With this choice we may use $\Sigma \simeq \omega$ and the solution \pref{exact12Schro} becomes
\begin{eqnarray}
\varrho_{12}(\tau) & \simeq & e^{-g^2 \CO \tau} \left\{ \varrho_{12}(0) e^{- i \omega \tau} + \varrho_{12}^{\ast}(0)\, \frac{g^2 \CO - i g^2 \DO}{\omega} \;\sin(\omega\tau) \right\} \,.\label{approx12Schro}
\end{eqnarray}
which is valid for times as late as $\omega \tau \sim \mathcal{O}( \omega / (g^2 \CBD) )$. Note in particular that \pref{g2constraint} implies that both $g^2 \CBD / \omega \ll 1$ and $g^2 \DBD / \omega \ll 1$, which are small but not necessarily as small as $\cO(g^2)$. The assumption \pref{g2constraint} also implies that the oscillations of \pref{approx12Schro} are underdamped, in that they are very fast compared to the solution's decay time.

Because $\CBD > 0$ (see \pref{CBDdef}) these solutions describe relaxation towards a late-time static state 
\begin{eqnarray}
\lim_{\tau \to \infty} \varrho(\tau) \ = \ \left[ \begin{matrix} \dfrac{1}{e^{2\pi \omega / H} + 1} & 0 \\ 0 &  \dfrac{1}{e^{ - 2\pi \omega / H} + 1} \end{matrix} \right] \ , 
\end{eqnarray}
which is clearly a thermal state -- of the form $e^{ - \beta \mathfrak{h}} / \mathrm{Tr}[ e^{ - \beta \mathfrak{h}}  ]$ -- with temperature given by the Gibbons-Hawking formula 
\be
  T = \frac{1}{\beta} = \frac{ H}{2\pi} \,. 
\ee
The relaxation is predicted to be exponential -- proportional to $e^{-\tau/\xi}$ -- with a characteristic time-scale that differs for the diagonal and off-diagonal components of $\boldsymbol{\varrho}$: $2\xi_{11} =  \xi_{12} =  \xi_\ssM$, where 
\begin{eqnarray}
\xi_{\ssM} := \frac{1}{g^2 \CBD} \ = \ \frac{4\pi^3}{g^2 H} \mathrm{sech}\left( \frac{\pi \omega}{H} \right) \left|  \Gamma\left( \frac{3}{4} + \frac{\nu}{2} +  \frac{i\omega}{2H} \right) \Gamma\left( \frac{3}{4} - \frac{\nu}{2} +  \frac{i\omega}{2H} \right) \right|^{-2} \,.\label{MarkovTimescale}
\end{eqnarray}
Asymptotic forms for this expression are given in subsequent sections.

\subsection{Validity of the Markovian limit}
\label{sec:MarkovianValidity}

We next quantify the domain of validity of the Markovian approximation as a function of the problem's parameters: $H$, $M_{\rm eff}$ (or $m$ and coupling $\xi$), $\omega$, and $g$. 

As emphasized above, the Markovian evolution eqs.~\pref{rho11three} and \pref{rho12three} rely on two conditions. First, the focus must be on late times, $\tau \gg \tau_c$, compared to the correlation (or fall-off) time defined by $\WBD(\tau)$; second, higher terms in the Taylor expansion of $\boldsymbol{\varrho}^\ssI(\tau - s)$ in powers of $s$ within the integrand should be negligible. Since $\mathrm{Re}[\WBD(s)]$ falls off exponentially with $s$ for $ s \gg \tau_c$, after integration the neglect of derivatives relies on $\tau_c \partial_\tau \boldsymbol{\varrho}$ being small compared with $\boldsymbol{\varrho}$. 

It is this second condition whose validity imposes constraints on the system parameters. To see how, recall that the Markovian solutions found above for $\varrho_{11}^{\ssI}(\tau)$ and $\varrho_{12}^{\ssI}(\tau)$ are linear combinations of exponentials of the form $ \exp \left[ \left(  - 1/\xi + i \Phi \right) \tau \right]$, where $1/\xi \sim g^2 \CBD$ and the phase $\Phi$ is either zero or $\omega$. Consistency therefore requires $| - 1/\xi + i \Phi| \ll 1/\tau_c$. Chasing through the definitions this turns out to imply (see Appendix \ref{App:Validity} or \cite{qubitpaper1} for details)
the conditions
\begin{eqnarray}
g^2 |\DBDp| \ll 1\ , \quad  g^2 |\CBDp| \ll 1 \ , \quad  \left|\frac{2\omega\CBDp}{\CBD} \right| \ll 1 \quad \mathrm{and} \quad \left|  \frac{2\omega\DBDp}{\CBD} \right| \ll 1 \ , \label{validity1}
\end{eqnarray}
where primes denote differentiation with respect to $\omega$. We explore the implications of these conditions in the non-degenerate special case when \pref{g2constraint} is satisfied, and so
\begin{eqnarray}
\left| \frac{g^2 \CBD}{\omega} \right| \ , \ \left| \frac{g^2 \DBD}{\omega} \right| \ll 1 \,. \label{validity3}
\end{eqnarray}
Similar conditions for validity of the Markovian approximation can also be made when \pref{g2constraint} is not satisfied but for simplicity we do not pursue these further here (see, however, \cite{qubitpaper1}). 

\begin{table}[t]
  \centering    
     \centerline{\begin{tabular}{ r|c|c|c|c| }
 \multicolumn{1}{r}{}
 & \multicolumn{1}{c}{$\CBD(\omega)$}
 & \multicolumn{1}{c}{$\underset{\ }{  \CBDp(\omega)}$}
 & \multicolumn{1}{c}{$\DBD(\omega) $}
 & \multicolumn{1}{c}{$\DBDp(\omega)$} \\
\cline{2-5}
$\frac{\omega}{H} \ll \frac{\kappa}{H} \simeq \frac{M_{\mathrm{eff}}^2}{3H^2} \ll 1$ & $ \stackrel{\ }{ \underset{\ }{ \dfrac{9H^5}{4 \pi^2 M_{\mathrm{eff}}^4} } } $ & $ - \dfrac{81 H^7 \omega}{2\pi^2 M_{\mathrm{eff}}^8} $ & $\dfrac{\omega \log(H \epsilon) }{2\pi^2} +  \dfrac{27H^6\omega}{4\pi^2M_{\mathrm{eff}}^6}$ &  $\dfrac{\log(H \epsilon)}{2\pi^2}   +  \dfrac{27H^6}{4\pi^2M_{\mathrm{eff}}^6}$ \\
\cline{2-5}
$\frac{\kappa}{H} \simeq \frac{M_{\mathrm{eff}}^2}{3H^2}  \ll \frac{\omega}{H} \ll 1$ & $\stackrel{\ }{  \underset{\ }{ \dfrac{H^3}{4\pi^2\omega^2} } } $ & $ - \dfrac{H^3}{2\pi^2\omega^3}$ &  $\dfrac{\omega\log(H \epsilon)}{2\pi^2} + \dfrac{3H^4}{4\pi^2\omega M_{\mathrm{eff}}^2}$ &  $\dfrac{\log(H \epsilon)}{2\pi^2}  - \dfrac{3H^4}{4\pi^2\omega^2 M_{\mathrm{eff}}^2}$  \\
\cline{2-5}
$\frac{M_{\mathrm{eff}}}{H} \ll  1 \ll \frac{\omega}{H}$ & $  \underset{\ }{ \dfrac{\omega}{4\pi} } $ & $\dfrac{1}{4\pi}$ & $\dfrac{\omega\log(e^{\gamma} \omega \epsilon)}{2\pi^2} $ &  $\stackrel{\ }{ \dfrac{\log(e^{\gamma + 1} \omega \epsilon)}{2\pi^2}  }$ \\
\cline{2-5}
$\frac{\omega}{H} \ll 1$ \& $\frac{M_{\mathrm{eff}}}{H} = \sqrt{2}$ & $ \stackrel{\ }{  \underset{\ }{ \dfrac{H}{4\pi^2} } } $ & $\dfrac{\omega}{6H}$ & $\dfrac{\omega\log(H \epsilon)}{2\pi^2} $ &  $ \stackrel{\ }{ \dfrac{ \log(H \epsilon)}{2\pi^2}}$ \\
\cline{2-5}
$1 \ll \frac{\omega}{H}$ \& $\frac{M_{\mathrm{eff}}}{H} = \sqrt{2}$ & $ \underset{\ }{ \dfrac{\omega}{4\pi} } $ & $\stackrel{\ }{ \dfrac{1}{4\pi} }$ &  $\dfrac{\omega\log(e^{\gamma} \omega \epsilon)}{2\pi^2} $ &  $ \stackrel{\ }{\dfrac{\log(e^{\gamma + 1} \omega \epsilon)}{2\pi^2} }$ \\
\cline{2-5}
$ \frac{\omega}{H} \ll  1 \ll\frac{M_{\mathrm{eff}}}{H}$ & $\stackrel{\ }{ \underset{\ }{ \dfrac{H}{\pi} e^{ - { \pi M_{\mathrm{eff}} }/{H} }  } } $ & $\dfrac{\pi \omega}{H} e^{- {\pi M_{\mathrm{eff}}}/{H}}$ &  $\dfrac{\omega\log(H \epsilon)}{2\pi^2} $ &  $\dfrac{\log(H \epsilon)}{2\pi^2} $ \\
\cline{2-5}
$1 \ll \frac{\omega}{H} \ll \frac{M_{\mathrm{eff}}}{H}$ & $ \stackrel{\ }{ \underset{\ }{ \dfrac{M_{\mathrm{eff}}}{4\pi} \; e^{{ \pi \omega}/{ H} }  e^{ - {\pi M_{\mathrm{eff}} }/{H} } } } $ & $\dfrac{M_{\mathrm{eff}}}{4H} \; e^{{ \pi \omega}/{ H} }  e^{ - {\pi M_{\mathrm{eff}} }/{H}  } $ &  $\dfrac{\omega\log(e^{\gamma} \omega \epsilon)}{2\pi^2} $ &  $ \stackrel{\ }{ \dfrac{\log(e^{\gamma + 1} \omega \epsilon)}{2\pi^2} }$ \\
\cline{2-5}
$1  \ll \frac{M_{\mathrm{eff}}}{H} \ll \frac{\omega}{H} $ & $ \stackrel{\ }{  \underset{\ }{ \dfrac{\omega}{4\pi} } } $ & $\dfrac{1}{4\pi}$ &  $\dfrac{\omega\log(e^{\gamma} \omega \epsilon)}{2\pi^2} $ &  $ \stackrel{\ }{ \dfrac{\log(e^{\gamma + 1} \omega \epsilon)}{2\pi^2} }$ \\
\cline{2-5}
\end{tabular} \quad \quad }
        \caption{Asymptotic behaviour of the functions $\CBD$, $\CBDp$, $\DBD$ and $\DBDp$ in various regimes differing in the relative sizes of $\omega$, $M_{\mathrm{eff}}$ and $H$ (see Appendix \ref{App:DBD} for the $\epsilon$-dependence for $\DBD$ and $\DBDp$). To illustrate the behaviour of the functions for intermediate values of mass, we provide the behaviour for the conformal scalar case with $M_{\mathrm{eff}} = \sqrt{2} H$. Here $\epsilon$ is the short-distance regulator for the divergence in $\DBD$. In the limit where $\omega/H \ll 1$ and $M_{\mathrm{eff}} / H \ll 1$ the functions become parametrically large (due to a singularity at $\omega = M_{\mathrm{eff}} = 0$) --- the sub-leading terms in these asymptotic series are either $\mathcal{O}(\omega/\kappa)$ or $\mathcal{O}(\kappa / \omega)$ ({\it e.g.} in the very first cell $\tfrac{9H^5}{4\pi^2M_{\mathrm{eff}}^4} \left[ 1 + \mathcal{O}( {\omega} / {\kappa} )\right]$) (see Appendix \ref{App:OrdLim}).  Note that in this regime the functions $\DBD$ and $\DBDp$ both acquire parametrically large terms which compete with the $\epsilon$-divergences.} \label{Functions1}
\end{table}

To explore the implications of \pref{validity1} for the parameters $\omega$, $H$, $M_{\rm eff}$ and $g$ it helps to have asymptotic expressions for the integrals $\CBD$ and $\DBD$ in various limits. These are summarized for convenience\footnote{These functions prove to be very singular in the limit where $\omega$ and $M_{\rm eff}$ both vanish, and so their asymptotic forms depend in an important way on the relative size of $\omega/H$ and $M_{\rm eff}/H$. See Appendix \ref{App:OrdLim} for details.} in Table \ref{Functions1}. 
Since the first two conditions in \pref{validity1} are proportional to $g$ they tend to be satisfied once $g$ is chosen deep enough in the perturbative regime. How deep depends on the relative sizes of the other parameters, as can be seen from Columns 2 and 4 of Table \ref{Functions1}. 

It is the second two conditions of \pref{validity1} that cannot be ensured simply by making $g$ small. The implications of the estimates in Table \ref{Functions1} for these two quantities are summarized in Table \ref{PossibleOrNot}. The final column indicates with checks or crosses whether these quantities can be small enough to allow a Markovian approximation. The first observation emerging from these tables is that a Markovian approximation necessarily requires $\omega / H \ll 1$. (That is,  the Markovian approximation can be satisfied only if $\omega$ is much smaller than the Gibbons-Hawking temperature. This result is intuitive because, if $\omega$ were much bigger than the temperature, interactions with the field (which is for the qubit effectively a thermal bath) become very inefficient at erasing the correlations whose absence underlies the qubit's thermalization and Markovian evolution.) Inspection of the last two rows of Table \ref{PossibleOrNot} shows that it is the condition $|\omega \CBDp /\CBD | \ll 1$ that generically fails if $\omega \gg H$.
 
\begin{table}[h]
  \centering    
     \centerline{\begin{tabular}{ r|c|c|c|c| }
 \multicolumn{1}{r}{}
 & \multicolumn{1}{c}{$\dfrac{2\omega\CBDp}{\CBD} $}
 & \multicolumn{1}{c}{$\underset{\ }{  \dfrac{2\omega\DBDp}{\CBD} }$}
 & \multicolumn{1}{c}{Markovian Limit?} \\
\cline{2-4}
$\frac{\omega}{H} \ll \frac{\kappa}{H} \simeq \frac{M_{\mathrm{eff}}^2}{3H^2} \ll 1$ & $ - \dfrac{36H^2\omega^2}{M_{\mathrm{eff}}^4} $ & $\stackrel{\ }{  \underset{\ }{ \dfrac{4\omega M_{\mathrm{eff}}^4 \log(H\epsilon) }{9H^5}+ \dfrac{6H\omega}{M_{\mathrm{eff}}^2} } }$ & $\cmark$ \\
\cline{2-4}
$\frac{\kappa}{H} \simeq \frac{M_{\mathrm{eff}}^2}{3H^2} \ll \frac{\omega}{H} \ll 1$ & $  - 4 $ & $\stackrel{\ }{  \underset{\ }{ \dfrac{4\omega^3\log(H\epsilon)}{H^3} - \dfrac{6H \omega}{M_{\mathrm{eff}}^2} } }$ & $\xmark$ \\
\cline{2-4}
$\frac{M_{\mathrm{eff}}}{H} \ll  1 \ll \frac{\omega}{H}$ & $2$ & $ \stackrel{\ }{  \underset{\ }{ \dfrac{4\log(e^{\gamma+1}\omega\epsilon)}{\pi} } } $ & $\xmark$ \\
\cline{2-4}
$\frac{\omega}{H} \ll 1$ \& $\frac{M_{\mathrm{eff}}}{H} = \sqrt{2}$ & $ \stackrel{\ }{  \underset{\ }{ \dfrac{4\pi^2\omega^2}{3H^2} } } $ & $\dfrac{2\omega\log(H\epsilon)}{H}$ & $\cmark\cmark$ \\
\cline{2-4}
$1 \ll \frac{\omega}{H}$ \& $\frac{M_{\mathrm{eff}}}{H} = \sqrt{2}$ & $2$ & $ \stackrel{\ }{  \underset{\ }{ \dfrac{4\log(e^{\gamma+1}\omega\epsilon)}{\pi} } } $ & $\xmark$ \\
\cline{2-4}
$ \frac{\omega}{H} \ll  1 \ll\frac{M_{\mathrm{eff}}}{H}$ & $ \stackrel{\ }{  \underset{\ }{ \dfrac{2\pi^2\omega^2}{H^2} } } $ & $\dfrac{\omega \log(H\epsilon)}{2\pi H} e^{+ {\pi M_{\mathrm{eff}} }/{ H } }$ & $\cmark$ \\
\cline{2-4}
$1 \ll \frac{\omega}{H} \ll \frac{M_{\mathrm{eff}}}{H}$ & $ \stackrel{\ }{  \underset{\ }{ \dfrac{2\pi\omega}{H} } } $ & $\dfrac{4 \omega \log(e^{\gamma+ 1} \omega \epsilon)}{\pi M_{\mathrm{eff}}} e^{+{\pi (M_{\mathrm{eff}} -\omega)}/{ H } }$ & $\xmark$ \\
\cline{2-4}
$1  \ll \frac{M_{\mathrm{eff}}}{H} \ll \frac{\omega}{H} $ & $2$ & $ \stackrel{\ }{  \underset{\ }{ \dfrac{4\log(e^{\gamma+1}\omega\epsilon)}{\pi} } } $ & $\xmark$ \\
\cline{2-4}
\end{tabular} \quad \quad }
        \caption{Asymptotic forms in different parameter regimes for the two $g$-independent conditions for the validity of the Markov approximation, as computed using the asymptotic forms provided in Table \ref{Functions1}. The third column indicates whether parameters exist that make the previous two terms small, without checking whether or not \pref{validity3} is also satisfied. Two checks indicate cases where $\omega/H$ is not required to be small in an $M_{\rm eff}/H$-dependent way. (Notice in particular that checks only appear in regimes where $\omega/ H$ is small).} \label{PossibleOrNot}
\end{table}

More generally, once $\omega \ll H$ is satisfied conditions \pref{validity1} can also be satisfied at sufficiently late times for any value of $M_{\mathrm{eff}} / H$ by choosing $\omega$ and $g$ to be sufficiently small. In practice, for some choices for $M_{\rm eff}$ the allowed value of $\omega$ can be small enough to push it to the point where \pref{validity3} is no longer true. In such a case the above formulae need not hold and a separate analysis must be done (along the lines given in \cite{qubitpaper1}). Two check marks in Table \ref{PossibleOrNot} indicate that the required value for $\omega/H$ is not parametrically small in a way that depends sensitively on $M_{\rm eff}/H$.

Table \ref{MarkovianTable1} displays the asymptotic behaviour of all of the conditions in both \pref{validity1} and \pref{validity3} to see how restrictive condition \pref{validity3} is for situations that have only one check mark in Table \ref{PossibleOrNot}. We note that satisfying all conditions in \pref{validity1} and \pref{validity3} can be done in all three cases by making $g$ and $\omega$ small enough. The allowed range for $\omega/H$ for which the Markovian limit applies becomes smaller and smaller for $M_{\mathrm{eff}} / H \ll 1$ and $\omega/H \lsim e^{-\pi M_{\rm eff}/H}$ is exponentially small when $M_{\rm eff} / H \gg 1$. 

By contrast no obstruction seems to arise to a Markovian limit when $M_{\rm eff}/ H$ is $\cO(1)$.

\begin{table}[t]
  \centering    
     \centerline{\begin{tabular}{ r|c|c|c|c|c|c| }
 \multicolumn{1}{r}{}
 & \multicolumn{1}{c}{$\frac{\omega}{H} \ll \frac{\kappa}{H} \simeq \frac{M_{\mathrm{eff}}^2}{3H^2} \ll 1$}
 & \multicolumn{1}{c}{$\underset{\ }{  \frac{\omega}{H} \ll 1\ \& \ \frac{M_{\mathrm{eff}}}{H} = \sqrt{2} }$ }
 & \multicolumn{1}{c}{$\frac{\omega}{H} \ll 1 \ll \frac{M_{\mathrm{eff}}}{H}$} \\
\cline{2-4}
$   g^2 |\CBDp(\omega) |$ & $ \stackrel{\ }{ \underset{\ }{\dfrac{81 g^2 H^7 \omega}{2 \pi^2 M_{\mathrm{eff}}^8} } } $ & $\dfrac{g^2 \omega}{6 H}$ & $\dfrac{\pi g^2 \omega}{H} e^{- {\pi M_{\mathrm{eff}}}/{H}}$ \\
\cline{2-4} 
$   g^2 | \DBDp(\omega) | $ & $ \stackrel{\ }{ \underset{\ }{\left| \dfrac{g^2 \log(H\epsilon)}{2\pi^2} + \dfrac{27g^2H^6}{4\pi^2M_{\mathrm{eff}}^6} \right| } } $ & $ \dfrac{g^2 |\log(H\epsilon)|}{2\pi^2}$ & $\dfrac{g^2 |\log(H\epsilon)|}{2\pi^2}$ \\
\cline{2-4} 
$  \left| \dfrac{2 \omega \CBDp}{\CBD} \right|  $ & $ \stackrel{\ }{ \underset{\ }{ \dfrac{36H^2\omega^2}{M_{\mathrm{eff}}^4}  } } $ & $\stackrel{\ }{ \underset{\ }{  \dfrac{4\pi^2\omega^2}{3H^2} } }$ & $  \dfrac{2\pi^2\omega^2}{H^2} $ \\
\cline{2-4} 
$  \left|  \dfrac{2\omega \DBDp }{\CBD} \right|$ & $ \stackrel{\ }{ \underset{\ }{ \left| \dfrac{g^2 \log(H\epsilon)}{2\pi^2} + \dfrac{6H\omega}{M_{\mathrm{eff}}^2} \right| } } $ & $\stackrel{\ }{ \underset{\ }{  \dfrac{2\omega\log(H\epsilon)}{H}  } }$ & $\dfrac{\pi\omega}{H} e^{+ {\pi M_{\mathrm{eff}}}/{H}}\left| \frac{\log(H\epsilon) }{2\pi^2}   \right|$ \\
\cline{2-4} 
$\left| \dfrac{g^2 \CBD(\omega)}{\omega} \right| $ & $ \stackrel{\ }{ \underset{\ }{\dfrac{9 g^2 H^5}{4\pi^2 \omega M_{\mathrm{eff}}^4 } } } $ & $\dfrac{g^2 H}{4\pi^2 \omega}$ & $\dfrac{g^2 H}{\pi \omega} e^{- {\pi M_{\mathrm{eff}}}/{H}}$ \\
\cline{2-4} 
$   \left| \dfrac{g^2 \DBD(\omega)}{\omega} \right|$ & $ \stackrel{\ }{ \underset{\ }{\left| \dfrac{g^2 \log(H\epsilon)}{2\pi^2} + \dfrac{27g^2H^6}{4\pi^2M_{\mathrm{eff}}^6} \right| } } $ & $ \dfrac{g^2 |\log(H\epsilon)|}{2\pi^2}$ & $\dfrac{g^2 |\log(H\epsilon)|}{2\pi^2}$ \\
\cline{2-4} 
\end{tabular} \quad \quad }
        \caption{Leading-order behaviour for the quantities appearing in conditions \pref{validity1} and \pref{validity3} for the three parameter regimes that receive checks in Table \ref{PossibleOrNot}. Although the first four rows here duplicate information in Table \ref{PossibleOrNot} about the size of terms appearin in \pref{validity1}, the last two rows compare this information with the asymptotic form for the combination appearing in condition \pref{validity3}.} \label{MarkovianTable1} 
\end{table}

We remark that in the regimes of both large and small $M_{\rm eff}/H$ inspection of Table \ref{Functions1} shows that the relaxation time-scale $\xi = (g^2 \CBD)^{-1}$ also becomes parametrically long. For large $M_{\rm eff}/H$ this seems due to the inefficiency of thermal interactions due to the Boltzmann suppression of excited field states. For small $M_{\rm eff}/H$ we think the system instead displays critical slowing down due to the large fluctuations that become possible as $M_{\rm eff} \to 0$. 
   
\subsection*{Positivity of the Markovian Limit}

A reality check for these approximation schemes is to verify that the Markovian limit preserves the hermiticity and normalization of the qubit density matrix. The argument given here to this end closely parallels the one given in \cite{qubitpaper1}.

The simplest way to establish preservation of hermiticity and normalization is to rewrite the Schr\"odinger-picture version of eqs.~\pref{rho11three} and \pref{rho12three} as a Lindblad equation \cite{Gorini:1975nb,Lindblad:1975ef,Gorini:1976cm,Redfield,DaviesOQS,Alicki,Kubo,Gardiner,Weiss,Breuer:2002pc,Rivas,Schaller}, which has the form
\begin{eqnarray}
\frac{\partial \boldsymbol{\varrho}(\tau)}{\partial \tau} & = & - i \big[ \mathfrak{h} + \sfrac{g^2}{2} ( \omega_1 + \DBD ) \boldsymbol{\sigma_{3}} , \boldsymbol{\varrho}(\tau) \big] + \sum_{j,k = 1}^{3} c_{jk} \left( \boldsymbol{F_{j}} \boldsymbol{\varrho}(\tau) \boldsymbol{F_{k}}^{\dagger} - \frac{1}{2} \left\{ \boldsymbol{F_{k}}^{\dagger} \boldsymbol{F_{j}}, \boldsymbol{\varrho}(\tau) \right\} \right) \ , \label{ExplicitLindblad}
\end{eqnarray}
for a basis of $2 \times 2$ matrices, $\boldsymbol{F_{j}} = \tfrac{1}{2} \boldsymbol{\sigma_{j}}$, and a collection of coefficients $\boldsymbol{c} = [c_{jk}]$ (called the Kossakowski matrix). The utility of writing the evolution in this form is that it is known to preserve hermiticity and normalization so long as the Kossakowski matrix is positive semi-definite. 

Eqs.~\pref{rho11three} and \pref{rho12three} can indeed be written in the form of \pref{ExplicitLindblad} provided that the Kossakowski matrix is given by
\begin{eqnarray}
\boldsymbol{c} & = & \left[ \begin{matrix} 4 g^2 \CO  & \  2g^2 (\DBD - i \SBD ) & 0 \\ 2g^2 (\DBD + i \SBD )  & 0 & 0 \\ 0& 0 & 0 \end{matrix} \right] \,, \label{KossaMatrix}
\end{eqnarray}
where the identity $\RBD = \CBD + \SBD$ from \pref{RrelatedCS} has been used. The eigenvalues of the Kossakowski matrix \pref{KossaMatrix}  are
\begin{eqnarray}
\lambda^{\boldsymbol{c}}_{1} & = & 0 \ , \nn\\
\lambda^{\boldsymbol{c}}_{2} & = & 2 g^2 ( \CBD + \sqrt{ \mathcal{C}_{\mathrm{BD}}^2 + \mathcal{S}_{\mathrm{BD}}^2 + \Delta_{\mathrm{BD}}^2 } \ )\ ,  \\
\mathrm{and} \quad \lambda^{\boldsymbol{c}}_{3} & = & 2 g^2 ( \CBD - \sqrt{ \mathcal{C}_{\mathrm{BD}}^2 + \mathcal{S}_{\mathrm{BD}}^2 + \Delta_{\mathrm{BD}}^2 } \ ) \ .\nn
\end{eqnarray}
At first sight the third eigenvalue is a problem because it is negative. We now briefly argue why this is not a problem: we show that within the domain of validity of the Markovian limit the eigenvalue $\lambda^c_3$ is actually consistent with zero. 

To see why this is so, consider for illustrative purposes the simplest regime, for which $\omega/H \ll 1$ and $M_{\rm eff}/H \simeq \sqrt2$. Table \ref{Functions1} shows that in this regime $\CBD \simeq H/(4\pi^2)$ is systematically larger than is $\DBD \simeq \omega \log(H \epsilon)/(2\pi^2)$. The same is also true for $\SBD = - \CBD \, \tanh(\pi \omega/H) \simeq -\omega/(4\pi) \ll \CBD$ -- see eq.~\pref{ratioSC}. As a consequence the two nonzero eigenvalues can be approximated by
\bea
  \lambda^{\boldsymbol{c}}_{2} & = & 2 g^2 ( \CBD + \sqrt{ \mathcal{C}_{\mathrm{BD}}^2 + \mathcal{S}_{\mathrm{BD}}^2 + \Delta_{\mathrm{BD}}^2 } \ )\  \simeq 2 g^2 \CBD \nn \\
\mathrm{and} \quad \lambda^{\boldsymbol{c}}_{3} & = & 2 g^2 ( \CBD - \sqrt{ \mathcal{C}_{\mathrm{BD}}^2 + \mathcal{S}_{\mathrm{BD}}^2 + \Delta_{\mathrm{BD}}^2 } \ )  \simeq - \frac{g^2 (\mathcal{S}_{\mathrm{BD}}^2 +  \Delta_{\mathrm{BD}}^2)}{\CBD}  \sim \cO(g^2 \omega^2/H) \,.
\eea
The negative eigenvalue is therefore comparable to terms that were neglected when deriving the Markovian limit, and is consistent with zero at the order being studied.  This also makes sense: unitarity and hermiticity are properties of the full theory's density matrix, and cannot be ruined by any approximation that accurately captures this underlying physics.

\section{Solved non-Markovian evolution when $M_{\rm eff} \ll H$}
\label{sec:NonMarkovian}

The previous section argues that the Markovian limit only exists when $\omega \ll H$, and is most robust in this regime when $M_{\rm eff} \sim H$. The Markovian approximation got harder to achieve in the critical-slowing-down regime where $M_{\mathrm{eff}} \ll H$.  

Fundamentally, Markovian evolution becomes harder to achieve for small $M_{\rm eff}$ because the Markovian approximation requires a hierarchy between the correlation time $\tau_c \to \kappa^{-1}$ of the Wightman function and the longer time $\xi$ over which the qubit evolves. As $M_{\rm eff}$ is decreased the correlation time $\kappa^{-1} \propto H/M_{\rm eff}^2$ grows larger and for generic choices of the parameters the qubit does not evolve slowly enough for the derivative expansion in the Nakajima-Zwanzig equation to be a good approximation.

In this section we focus on this $M_{\mathrm{eff}} \ll H$ regime, and show how to solve for the system's evolution without requiring access to the Markovian limit. This allows us to track explicitly the memory effects coming from the very long tail of the Wightman function. By doing so, we widen the regime of parameter space for which solutions for the late-time evolution of $\boldsymbol{\varrho}(\tau)$ are explicitly known. 

\subsection{Non-Markovian evolution}
\label{sec:NonMarkovianSetUp}

We first recall the form of the subleading tail $f(\tau)$ defined in \pref{subleading},
\begin{eqnarray}
f(\tau) & = & \WBD(\tau) - \mathcal{W}_0 \,e^{-\kappa\tau} \label{tail2}
\end{eqnarray}
for $\tau > 0$, with $\mathcal{W}_0$ and $\kappa$ defined in \pref{W0def}. In the $M_{\mathrm{eff}}/H \ll 1$ regime of interest the leading-order normalization $\mathcal{W}_0$ and time-scale $1/\kappa$ admit the expansions
\begin{eqnarray}
\frac{1}{\kappa} & \simeq & \frac{3H}{M_{\mathrm{eff}}^2} - \frac{1}{3H} + \ldots \ldots \ , \label{kappaExpansion2} \\
\mathcal{W}_0 \ = \ \mathrm{Re}[\mathcal{W}_0] + i \mathrm{Im}[\mathcal{W}_0] & \simeq & \bigg[ \frac{3H^4}{8\pi^2 M_{\mathrm{eff}}^2} - \frac{(7-6\ln 2)H^2}{24\pi^2} + \ldots \bigg] + i \bigg[ - \frac{H^2}{8\pi^2} + \ldots \bigg] \ . \quad \quad \label{WoExpansion}
\end{eqnarray}

In particular the sub-leading tail $f(\tau) \sim \mathcal{O}(e^{-H\tau})$ falls off exponentially on the {\it much} faster time-scale $1/H$ than does $\WBD(\tau)$. Our strategy for solving the Nakajima-Zwanzig equation is to include the memory effects contained in $\cW_0 \, e^{-\kappa \tau}$ explicitly and use the Taylor-expansion argument only for integrals containing the much more quickly-falling function $f(\tau)$.

In doing so, it will be convenient to define two integral transforms (closely related to $\CBD$ and $\DBD$) which will prove useful in the critical slowing-down regime. In analogy to \pref{CBDdef} and \pref{DBDdef} we define
\begin{eqnarray}
\tCBD & := & 2 \int_0^\infty \exd s\ \mathrm{Re}[f(s)] \cos(\omega s) \ = \ \CBD - \frac{2 \RWo \kappa}{\kappa^2 + \omega^2} \ , \label{tCBDdef}\\
\tDBD & := & 2 \int_0^\infty \exd s\ \mathrm{Re}[f(s)] \sin(\omega s) \ = \ \DBD - \frac{2 \RWo \omega}{\kappa^2 + \omega^2} \ . \label{tDBDdef}
\end{eqnarray}
As can be seen in Table \ref{Functions2} these functions along with their $\omega$-derivatives, $\tCBDp$ and $\tDBDp$, are notably better behaved in the $M_{\mathrm{eff}} / H \ll 1$ regime (as opposed to their counterparts in Table \ref{Functions1} which have singularities at $M_{\mathrm{eff}} = \omega = 0$).

\begin{table}[h]
  \centering    
     \centerline{\begin{tabular}{ r|c|c|c|c| }
 \multicolumn{1}{r}{}
 & \multicolumn{1}{c}{$\tCBD(\omega) $}
 & \multicolumn{1}{c}{$\underset{\ }{  \tCBDp(\omega)}$}
 & \multicolumn{1}{c}{$\tDBD(\omega)$}
 & \multicolumn{1}{c}{$\tDBDp(\omega)$} \\
\cline{2-5}
$\frac{\omega}{H} \ll \frac{\kappa}{H} \simeq \frac{M_{\mathrm{eff}}^2}{3H^2} \ll 1$ & $ \stackrel{\ }{ \underset{\ }{ \dfrac{(\pi^2 + 3)H}{12 \pi^2} } } $ & $\dfrac{(15-\pi^2)\omega}{90H}$ & $\dfrac{\omega \log(H \epsilon) }{2\pi^2}$ &  $\dfrac{\log(H \epsilon)}{2\pi^2}  $ \\
\cline{2-5}
$\frac{\kappa}{H} \simeq \frac{M_{\mathrm{eff}}^2}{3H^2} \ll \frac{\omega}{H} \ll 1$ & $\stackrel{\ }{  \underset{\ }{ \dfrac{(\pi^2 + 3)H}{12 \pi^2}  } } $ & $\dfrac{(15-\pi^2)\omega}{90H}$ &  $\dfrac{\omega\log(H \epsilon)}{2\pi^2}$ &  $\dfrac{\log(H \epsilon)}{2\pi^2}$  \\
\cline{2-5}
$\frac{M_{\mathrm{eff}}}{H} \ll  1 \ll \frac{\omega}{H}$ & $  \underset{\ }{ \dfrac{\omega}{4\pi} } $ & $\dfrac{1}{4\pi}$ & $\dfrac{\omega\log(e^{\gamma} \omega \epsilon)}{2\pi^2} $ &  $\stackrel{\ }{ \dfrac{\log(e^{\gamma + 1} \omega \epsilon)}{2\pi^2}  }$ \\
\cline{2-5}
\end{tabular} \quad \quad }
        \caption{Leading-order behaviour for the functions $\tCBD$, $\tCBDp$, $\tDBD$ and $\tDBDp$ in various regimes of relative sizes of $\omega$, $M_{\mathrm{eff}}$ and $H$, where $M_{\mathrm{eff}} / H \ll 1$. These functions are markedly better behaved near $(M_{\mathrm{eff}}, \omega) = (0,0)$ than the functions listed in Table \ref{Functions1}.} \label{Functions2}
\end{table}

The first step is to rewrite the Nakajima-Zwanzig equation in terms of $f(\tau)$ by explicitly expanding $\WBD(\tau) = \mathcal{W}_0 e^{- \kappa \tau} + f(\tau)$ in \pref{INTpictureNZ}, so that
\begin{eqnarray}
\frac{\partial \boldsymbol{\varrho}^\ssI (\tau)}{\partial \tau} & \simeq & g^2 \int_0^\tau \exd s\ \bigg( \left[ \mathcal{W}_0 e^{- \kappa (\tau - s) } + f(\tau - s) \right] \big[ \mathfrak{m}^\ssI(s) \, \boldsymbol{\varrho}^\ssI (s), \mathfrak{m}^\ssI(\tau) \big] \\
& \ & \quad \quad \quad \quad \quad \quad + \left[ \mathcal{W}^{\ast}_0 e^{- \kappa (\tau - s) } + f(\tau - s)^{\ast} \right]  \big[ \mathfrak{m}^\ssI(\tau) , \boldsymbol{\varrho}^\ssI (s)\, \mathfrak{m}^\ssI(s)  \big] \bigg) + i \left[ \frac{g^2 \DBD}{2} \boldsymbol{\sigma_{3}} , \boldsymbol{\varrho}^{\ssI}(\tau) \right] \,, \notag 
\end{eqnarray}
where we (as before) set the counter-term $\omega_1 = - \DBD$. In components (again using $\varrho^{\ssI}_{22}  = 1 - \varrho^{\ssI}_{11}$ and $\varrho^{\ssI}_{21} = \varrho^{\ssI\ast}_{12}$ to eliminate $\varrho_{22}$ and $\varrho_{21}$) we retrieve the equations of motion
\begin{eqnarray}
\frac{\partial \varrho^{\ssI}_{11}}{\partial \tau} & = & 2 g^2 \int\limits_{0}^{\tau} \exd s \; \left[ \mathrm{Re}[f(s)] \cos(\omega s) + \mathrm{Im}[f(s)] \sin(\omega s)  \right] + 2 g^2 \int\limits_{0}^{\tau} \exd s \; e^{-\kappa s} \left[ \RWo \cos(\omega s) + \mathrm{Im}[\mathcal{W}_0] \sin(\omega s)  \right] \notag \\
&\ & \quad - 4 g^2 \int_0^{\tau} \exd s\ \mathrm{Re}[f(s) ] \, \cos(\omega s) \varrho^{\ssI}_{11}(\tau - s) - 4 g^2 \RWo \int_0^{\tau} \exd s\ e^{- \kappa s} \, \cos(\omega s) \varrho^{\ssI}_{11}(\tau - s)  \, , \label{rho111NM} \\
\frac{\partial \varrho^{\ssI}_{12}}{\partial \tau} & = & i g^2 \DBD \; \varrho^{\ssI}_{12}(\tau) - 2 g^2 \int_0^\tau \exd s \ \mathrm{Re}[f(s)] e^{+ i \omega s} \varrho_{12}^{\ssI}(\tau - s)  + 2 g^2 e^{+ 2 i \omega \tau} \int_0^\tau \exd s \ \mathrm{Re}[f(s)] e^{- i \omega s} \varrho_{12}^{\ssI\ast}(\tau - s) \notag \\
&\ & \quad - 2 g^2 \RWo \int_0^\tau \exd s \ e^{ ( - \kappa + i \omega ) s} \varrho_{12}^{\ssI}(\tau - s) + 2 g^2 \RWo e^{+ 2 i \omega \tau} \int_0^\tau \exd s \ e^{ ( - \kappa - i \omega ) s} \varrho_{12}^{\ssI\ast}(\tau - s) \; . \quad \quad \quad \label{rho121NM}
\end{eqnarray}
So far these are exactly the same equations as  \pref{rho111}-\pref{rho121}, just written in a slightly different manner.  

The new step is to perform the same Taylor expansion of $\boldsymbol{\varrho}(\tau -s )$ as before, but only for those integrals involving $f(s)$. For these integrals this Taylor expansion is a much better approximation because the function $f(\tau)$ is correlated over much shorter times, of order $H^{-1}$. Keeping only the leading term in these Taylor expansions then implies 
\begin{eqnarray}
\frac{\partial \varrho^{\ssI}_{11}}{\partial \tau} & \simeq & 2 g^2 \int\limits_{0}^{\tau} \exd s \; \left[ \mathrm{Re}[f(s)] \cos(\omega s) + \mathrm{Im}[f(s)] \sin(\omega s)  \right] + 2 g^2 \int\limits_{0}^{\tau} \exd s \; e^{-\kappa s} \left[ \RWo \cos(\omega s) + \mathrm{Im}[\mathcal{W}_0] \sin(\omega s)  \right] \notag \\
&\ & \quad - 4 g^2 \int_0^{\tau} \exd s\ \mathrm{Re}[f(s) ] \, \cos(\omega s) \varrho^{\ssI}_{11}(\tau) - 4 g^2 \RWo \int_0^{\tau} \exd s\ e^{- \kappa s} \, \cos(\omega s) \varrho^{\ssI}_{11}(\tau - s)  \, , \label{rho111NM2} \\
\frac{\partial \varrho^{\ssI}_{12}}{\partial \tau} & \simeq & i g^2 \DBD \; \varrho^{\ssI}_{12}(\tau) - 2 g^2 \int_0^\tau \exd s \ \mathrm{Re}[f(s)] e^{+ i \omega s} \varrho_{12}^{\ssI}(\tau)  + 2 g^2 e^{+ 2 i \omega \tau} \int_0^\tau \exd s \ \mathrm{Re}[f(s)] e^{- i \omega s} \varrho_{12}^{\ssI\ast}(\tau) \notag \\
&\ & \quad - 2 g^2 \RWo \int_0^\tau \exd s \ e^{ ( - \kappa + i \omega ) s} \varrho_{12}^{\ssI}(\tau - s) + 2 g^2 \RWo e^{+ 2 i \omega \tau} \int_0^\tau \exd s \ e^{ ( - \kappa - i \omega ) s} \varrho_{12}^{\ssI\ast}(\tau - s) \; , \quad \quad \quad \label{rho121NM2}
\end{eqnarray}
where higher derivatives in the Taylor series are now order $(H^{-1} \partial_\tau)^n \varrho_{ij}$ and have been dropped. Note that the above equations explicitly track convolutions of $\boldsymbol{\varrho}^{\ssI}$ with the slow correlations $\sim e^{-\kappa s}$.

Again replacing the upper limits of integration by $\simeq \infty$, but only for the integrals involving $f(s)$ then leads to
\begin{eqnarray}
\frac{\partial \varrho^{\ssI}_{11}}{\partial \tau} & \simeq & g^2 \RBD - 2 g^2 \, \frac{\RWo \kappa + \mathrm{Im}[\mathcal{W}_0] \omega}{\kappa^2 + \omega^2} \, e^{-\kappa \tau} \cos(\omega \tau)  - 2 g^2 \,\frac{\RWo \omega - \mathrm{Im}[\mathcal{W}_0] \kappa}{\kappa^2 + \omega^2} \,e^{-\kappa \tau} \sin(\omega \tau) \label{rho111NM4} \nn\\
&\ & \quad  \quad  \quad \quad  \quad  \quad - 2 g^2 \tCBD \varrho^{\ssI}_{11}(\tau) - 4 g^2 \RWo \int_0^{\tau} \exd s\ e^{- \kappa s} \, \cos(\omega s) \varrho^{\ssI}_{11}(\tau - s)  \, ,   \\
\frac{\partial \varrho^{\ssI}_{12}}{\partial \tau} & \simeq & - g^2 \left( \tCBD - i \frac{2\RWo \omega}{\kappa^2 + \omega^2} \right) \varrho_{12}^{\ssI}(\tau)  + g^2 e^{+ 2 i \omega \tau} ( \tCBD - i \tDBD ) \varrho_{12}^{\ssI\ast}(\tau) \label{rho121NM4m} \\
&\ & \quad - 2 g^2 \RWo \int_0^\tau \exd s \ e^{ ( - \kappa + i \omega ) s} \varrho_{12}^{\ssI}(\tau - s) + 2 g^2 \RWo e^{+ 2 i \omega \tau} \int_0^\tau \exd s \ e^{ ( - \kappa - i \omega ) s} \varrho_{12}^{\ssI\ast}(\tau - s) \; , \notag \quad \quad
\end{eqnarray}
which uses the definitions \pref{tCBDdef} and \pref{tDBDdef} and, as before, $\RBD$ is given by \pref{RBDdef}. We next solve these equations explicitly without the need for further approximations. We return at the end to quantify the domain of validity of dropping the subdominant terms in the Taylor expansion. 

\subsection{The non-Markovian off-diagonal equation}
\label{sec:NMOffDiagonal}

It turns out that it is simpler to first solve the off-diagonal equation of motion. Using the relation $\varrho_{12}(\tau) = e^{- i \omega \tau} \varrho^{\ssI}_{12}(\tau)$, we transform \pref{rho121NM4m} to the Schr\"odinger-picture, which is easier to solve:
\begin{eqnarray}
\frac{\partial \varrho_{12}}{\partial \tau} & \simeq & \left( - g^2 \tCBD - i \omega + i \frac{2 g^2 \RWo \omega }{\kappa^2 + \omega^2} \right) \varrho_{12}(\tau) + ( g^2 \tCBD - i g^2 \tDBD ) \varrho_{12}^{\ast}(\tau) \label{rho121NM4} \\
&\ & \quad \quad \quad \quad - 2 g^2 \RWo \int_0^\tau \exd s \ e^{ - \kappa s} \bigg[ \varrho_{12}(\tau - s) - \varrho_{12}^{\ast}(\tau - s) \bigg] \notag
\end{eqnarray}
We solve this equation by converting it to a system of first-order linear differential equations. To this end define the three functions
\begin{eqnarray}
x_1 (\tau) &:=& \varrho_{12}(\tau) \\
x_2 (\tau) &:=& \varrho_{12}^{\ast}(\tau) \\
x_3(\tau) &:=& \int_0^\tau \exd s \ e^{ - \kappa s} \bigg[ \varrho_{12}(\tau - s) - \varrho_{12}^{\ast}(\tau - s) \bigg] \ . \label{x3def}
\end{eqnarray}
We regard $x_2 (\tau)$ and $x_3(\tau)$ only as auxiliary functons, and our real interest is in the solution for $x_1 (\tau) = \varrho_{12}(\tau)$. 

We now show that the functions $x_i(\tau)$ satisfy a set of coupled first-order differential equations that are easily solved. In terms of these new variables the equation of motion \pref{rho121NM4} can be recast as
\begin{eqnarray}
\frac{\exd {x}_{1}}{\exd \tau} = \left( - g^2 \tCBD - i \omega + i \frac{2 g^2 \RWo \omega }{\kappa^2 + \omega^2} \right) x_1(\tau) + ( g^2 \tCBD - i g^2 \tDBD ) x_2(\tau) - 2 g^2 \RWo x_{3}(\tau) \ . \quad \quad
\end{eqnarray}
Taking the complex conjugate of this gives a second evolution equation
\begin{eqnarray}
\frac{\exd {x}_{2}}{\exd \tau} = ( g^2 \tCBD + i g^2 \tDBD ) x_1(\tau) +  \left( - g^2 \tCBD + i \omega - i \frac{2 g^2 \RWo \omega }{\kappa^2 + \omega^2} \right) x_2(\tau)  + 2 g^2 \RWo x_{3}(\tau) \ . \quad \quad
\end{eqnarray}
Using the property $\int_0^t \exd s \; g(s) h(\tau - s) = \int_0^t \exd s \; g(\tau- s) h(s)$ of the convolution, the definition \pref{x3def} of $x_{3}(\tau)$ may be directly differentiated to yield the third equation,
\begin{eqnarray}
\frac{\exd {x}_{3}}{\exd \tau}= x_{1}(\tau) - x_{2}(\tau) - \kappa x_{3}(\tau)
\end{eqnarray}

In vector notation, defining $\mathbf{x}(\tau) = \big( x_{1}(\tau), x_{2}(\tau), x_{3}(\tau) \big)$, the above three equations close as a system of differential equations of the form
\begin{eqnarray}
\frac{\exd {\mathbf{x}}}{\exd \tau}  & = & \mathbb{O} \mathbf{x}(\tau) \label{vecO}
\end{eqnarray}
with matrix 
\begin{eqnarray} \label{xODE}
\mathbb{O} := \left[ \begin{matrix}  - g^2 \tCBD - i \omega + i \frac{2 g^2 \RWo \omega}{\kappa^2 + \omega^2} & g^2 \tCBD - i g^2 \tDBD & - 2 g^2 \RWo \\
g^2 \tCBD + i g^2 \tDBD &  - g^2 \tCBD + i \omega - i \frac{2 g^2 \RWo \omega }{\kappa^2 + \omega^2} & 2 g^2 \RWo \\
1 & - 1 & - \kappa \end{matrix} \right] 
\end{eqnarray}
and initial condition
\begin{eqnarray}
\mathbf{x}(0) = \left[ \begin{matrix} \varrho_{12}(0) \\ \varrho_{12}^{\ast}(0) \\ 0 \end{matrix} \right] \, .
\end{eqnarray}
Using \pref{tDBDdef} to relate $\tDBD$ to $\DBD$, the determinant of the matrix $\mathbb{O}$ is
\begin{eqnarray}
\det\mathbb{O}  \ = \ - \kappa \omega^2 \left( 1 - \frac{4g^2 \RWo}{\kappa^2 + \omega^2} +\frac{4 g^2 \RWo}{\kappa^2 + \omega^2} \left( \frac{g^2 \DBD}{\omega} \right) - \frac{g^4\Delta_{\mathrm{BD}}^2}{\omega^2} \right) \ \neq \ 0 \ ,
\end{eqnarray}
which is non-zero in the regime of interest that follows (for which to good approximation $\det\mathbb{O} \simeq - \kappa \omega^2$). Because $\mathbb{O}$ is invertible the unique static solution for the system is 
\be
    \mathbf{x}^{\mathrm{static}} = \mathbf{0} \,. 
 \ee
The general solution to \pref{xODE} is given by
\begin{eqnarray}
\mathbf{x}(\tau) = e^{\tau\mathbb{O}} \mathbf{x}(0) \label{Oic}
\end{eqnarray}
where specifically solving the first component $y_{1}(\tau) = \varrho_{12}(\tau)$ is of interest. Although the above matrix exponential can be exactly computed, the resulting solution is extremely unwieldy --- we will instead compute the above solution {\it perturbatively} using standard methods \cite{Hinch,Nayfeh}. The above equation of motion can be used to probe various regimes of small $g$, $\omega/H$, $M_{\mathrm{eff}} / H$: we demonstrate it's utility here by perturbing the solution in $g\ll 1$.

The dynamics of the matrix exponential are governed by the eigenvalues $\lambda^{\mathbb{O}}$ of $\mathbb{O}$, which can be computed from its characteristic polynomial
\begin{eqnarray}
0 & = & \big( \lambda^{\mathbb{O}} \big)^3 + \kappa \left( 1 + \frac{2 g^2 \tCBD}{\kappa} \right)  \big( \lambda^{\mathbb{O}} \big)^2 + \omega^2 \left( 1 + \frac{2 g^2 \kappa \CBD}{\omega^2} + \frac{g^2 \DBD}{\omega} \left[ \frac{4 g^2 \RWo}{\kappa^2 + \omega^2} - \frac{g^2 \DBD}{\omega} \right] \right) \lambda^{\mathbb{O}} \notag \\
& \ & \quad \quad \quad \quad \quad \quad \quad \quad + \kappa \omega^2 \left( 1 - \frac{4g^2 \RWo}{\kappa^2 + \omega^2} +\frac{4 g^2 \RWo}{\kappa^2 + \omega^2} \left( \frac{g^2 \DBD}{\omega} \right) - \frac{g^4\Delta_{\mathrm{BD}}^2}{\omega^2} \right) \ . \label{Ochar}
\end{eqnarray}
where $\tCBD$ and $\tDBD$ have been related to $\CBD$ and $\DBD$ using \pref{tCBDdef}-\pref{tDBDdef} (although a single factor of $\tCBD$ has been left in the first line of the above). If the qubit-field coupling were to vanish, the above equation becomes
\begin{eqnarray}
0 & = & \big( \lambda^{\mathbb{O}} \big)^3 + \kappa \big( \lambda^{\mathbb{O}} \big)^2 + \omega^2 \lambda^{\mathbb{O}} + \kappa \omega^2 \quad \quad \quad (\mathrm{if}\ g=0)
\end{eqnarray}
which yields `free' eigenvalues $\pm i \omega, -\kappa$. We here seek a solution slightly deviated away from the above free equation, as a perturbation in $g\ll 1$. In doing so, several dimensionless quantities in \pref{Ochar} need to be bounded in order to control the proceeding expansion for the eigenvalues. It turns out that sufficient conditions to bound the perturbation series are
\begin{eqnarray}
\frac{g^2 \kappa \CBD}{\omega^2} & \ll & 1 \ , \label{12bound1} \\
\frac{g^2 \DBD}{\omega} & \ll & 1 \ , \label{12bound2} \\
\frac{g^2 \CBD}{\kappa} \simeq \frac{2g^2\RWo}{\kappa^2 + \omega^2} & \ll & 1 \ . \label{12bound3}  
\end{eqnarray}
In particular in the third bound, to leading order uses the fact that
\begin{eqnarray}
\frac{g^2 \CBD}{\kappa} \simeq \frac{2g^2\RWo}{\kappa^2 + \omega^2} \ ,
\end{eqnarray}
which may be easily verified in the regimes of interest (either when $\kappa \gg \omega$ or $\omega \gg \kappa$). In particular, the bounds ${g^2 \DBD}/{\omega} \ll 1$ and ${2g^2\RWo}/{(\kappa^2 + \omega^2)}\ll 1$ ensure that the $\mathcal{O}(g^4)$ terms in \pref{Ochar} are negligibly small. The remaining $\mathcal{O}(g^2)$ terms are therefore small given that ${g^2 \kappa \CBD}/{\omega^2} \ll 1$ also holds.\footnote{Note that bound \pref{12bound3} necessarily implies that $2g^2\tCBD/\kappa \ll 1$ in the first line of \pref{Ochar}, as can be easily verified using Table \ref{Functions2}.}

Satisfying these bounds and perturbing the eigenvalues $\lambda_{\mathbb{O}}$ in $g$ yields 
\begin{eqnarray}
\lambda_{1}^{\mathbb{O}} &  \simeq & - i \omega \left[ 1 + \ldots  \right] - g^2 \CBD \left[ 1 + \ldots \right]   \, , \\
\lambda_{2}^{\mathbb{O}} & = & \lambda_{1}^{\mathbb{O}\ast}  \ , \label{Oeigenvalues} \\
\lambda_{3}^{\mathbb{O}} & \simeq & - \kappa \left[ 1 - \frac{4g^2\RWo}{\kappa^2 + \omega^2} + \ldots \right] \ , 
\end{eqnarray}
where the ellipsis denote negligibly small $\mathcal{O}(g^4)$ combinations of the dimensionless variables in \pref{12bound1}-\pref{12bound3}. Clearly the real part of each of the three eigenvalues are negative which means that the solution sinks towards the steady state $\mathbf{x}^{\mathrm{static}} = \mathbf{0}$. The first two eigenvalues clearly capture information relating to the Markovian approximation (with the Markovian time-scale $\xi_{\ssM}$), while the latter eigenvalue corresponds to a novel time-scale $\simeq 1/\kappa + \mathcal{O}(g^2)$.

The right eigenvectors $\mathbf{r}^{\mathbb{O}}$ corresponding to the eigenvalues \pref{Oeigenvalues} are 
\begin{eqnarray}
\mathbf{r}_{1}^{\mathbb{O}} \simeq \left[ \begin{matrix} 2 (\omega + i \kappa) - g^2 \frac{\CBD(\kappa + i \omega) + \DBD ( \omega + i \kappa  ) }{\omega}  \\ - g^2 \frac{(\CBD + i \DBD)(\kappa - i \omega)}{\omega} \\ 1 \end{matrix} \right] \, , \ \ \  \mathbf{r}_{2}^{\mathbb{O}} \simeq \left[ \begin{matrix} - g^2 \frac{(\CBD - i \DBD)(\kappa + i \omega)}{\omega} \\ 2 (\omega - i \kappa) - g^2 \ \frac{\CBD( \kappa - i \omega) + (\omega - i \kappa) \DBD}{\omega} \\ 1 \end{matrix} \right] \, , \quad \quad \\
\mathbf{r}_{3}^{\mathbb{O}} \simeq \left[ \begin{matrix} g^2 \frac{4 i \RWo}{\kappa - i \omega}  \\  - g^2 \frac{4 i \RWo}{\kappa + i \omega} \\ 1 \end{matrix} \right] \ , \notag \quad \quad \quad \quad \quad \quad  \quad \quad \quad \quad \quad \quad  \quad \quad \quad \quad \quad 
\end{eqnarray}
which can be used to solve for the solution. An ansatz of the form $\mathbf{x}(\tau) = \sum_{j} c_{j} \mathbf{r}_{j}^{\mathbb{O}}e^{ \lambda_{j}^{\mathbb{O}} \tau }$ explicitly satisfies \pref{vecO}, with $c_{j} $ being coefficients that can be perturbatively solved for using the initial condition \pref{Oic}. After a  straightforward computation, the solution for $x_{1}(\tau) = \varrho_{12}(\tau)$, which then transformed to the interaction-picture is found to be 
\begin{eqnarray}
\varrho^{\ssI}_{12}(\tau) &\simeq& \varrho_{12}(0) \left[ 1 + \frac{2 g^2 \RWo}{(\kappa - i \omega)^2} \right] e^{- g^2 \CBD \tau} \label{NMsol12} \\
& \ & \quad \quad \quad  + \varrho_{12}^{\ast}(0) \left[ \frac{g^2 \DBD + i g^2 \CBD }{2\omega} (1 - e^{+ 2 i \omega \tau} ) - \frac{2g^2 \RWo}{\kappa^2 + \omega^2} \right] e^{- g^2 \CBD \tau} \notag \\
& \ & \quad \quad \quad \quad \quad \quad \quad \quad+ \ \left[ - \varrho_{12}(0) \frac{2 g^2 \RWo}{(\kappa - i \omega)^2}  + \varrho_{12}^{\ast}(0) \frac{2g^2 \RWo}{\kappa^2 + \omega^2} \right] e^{+ i \omega \tau} \, e^{ - \kappa \left[ 1 - \tfrac{4 g^2 \RWo}{\kappa^2 + \omega^2} \right] \tau} \notag \ .
\end{eqnarray}
This solution contains much more information than the earlier Markovian solution {\it c.f.\ }\pref{approx12int} (or \pref{approx12Schro} in the Schr\"odinger-picture). Note that the bounds \pref{12bound1}-\pref{12bound3} ensure all the $\mathcal{O}(g^2)$ terms are small, as expected (in particular $g^2 \CBD / \omega \ll 1$ follows from \pref{12bound1} and \pref{12bound3} in both regimes $\omega \ll \kappa$ and $\kappa \ll \omega$ of interest). 

\subsection*{Validity relations for the off-diagonal equation}

Having established the above non-Markovian solution, the validity conditions for dropping derivatives in the expansion $\varrho_{12}^{\ssI}(\tau - s) \simeq \varrho_{12}^{\ssI}(\tau)$ must be derived next. The interaction-picture solution \pref{NMsol12} for $\varrho_{12}^{\ssI}(\tau)$ is (as in \S\ref{sec:MarkovianValidity} before) a linear combination of factors $\exp\big( [ - {1}/{\xi} + i \Phi] \tau \big)$ where the time-scales and phases are $(\xi,\Phi) \in \left\{ (\xi_\ssM, 0), (\xi_{\ssM},2\omega), (1/\kappa,\omega) \right\}$ (ignoring the small corrections to the time-scale $1/\kappa$ for the time being). For this reason, the validity relations are
\begin{eqnarray}
|\tCBD| \gg \left| \frac{\tDBDp}{\xi} - \tCBDp \Phi \right| \quad \quad \mathrm{and} \quad \quad |\tCBD| \gg \left| \frac{\tCBDp}{\xi} + \tDBDp \Phi \right| \ , \label{GeneralValidity2}
\end{eqnarray}
which are almost the same as \pref{GeneralValidity} derived in the Markovian limit in \S\ref{sec:MarkovianValidity} (see also \cite{qubitpaper1}). The derivation of the Markovian validity relations relies only on the exponential form of the ansatz $\exp\big( [ - {1}/{\xi} + i \Phi] \tau \big)$, which is why it applies here as well. The main difference in the above is that functions involved are the integral transforms associated with the correlation function $f$ (rather than $\WBD$ as in the Markovian limit).

All that remains is to insert $(\xi,\Phi) \in \left\{ (\xi_\ssM, 0), (\xi_{\ssM},2\omega), (1/\kappa,\omega) \right\}$ into the relation \pref{GeneralValidity2} (where we recall that the Markovian time-scale is defined by $1 / \xi_{\ssM} = g^2 \CBD$). Using $(\xi_\ssM, 0)$ and $(\xi_{\ssM},2\omega)$ leads\footnote{Given $|a|\ll 1$ and $|a-2b| \ll 1$ then $|b| \ll 1$ has been used here.} to the bounds
\begin{eqnarray}
\left| \frac{g^2 \CBD \tDBDp}{\tCBD} \right| \ll 1\ , \ \ \ \left| \frac{g^2 \CBD \tCBDp}{\tCBD} \right| \ll 1\ , \ \ \ \left| \frac{\omega \tDBDp}{\tCBD} \right| \ll 1 \ , \ \ \ \left| \frac{\omega \tCBDp}{\tCBD} \right| \ll 1 \ . \label{NM12bounds1}
\end{eqnarray}
And using $(1/\kappa,\omega)$ (along with the latter two bounds in \pref{NM12bounds1} above) leads to the bounds
\begin{eqnarray}
\left| \frac{\kappa \tDBDp}{2 \tCBD} \right| \ll 1\ , \ \ \ \left| \frac{\kappa \tCBDp}{2 \tCBD} \right| \ll 1 \ . \label{NM12bounds2}
\end{eqnarray}
All six of these bounds need to be satisfied in order for the above non-Markovian solution to be valid. Note that once again these relations only allow $\omega/H \ll 1$ (as in the Markovian case). In Table \ref{NMBounds1}, the functions in the above six bounds are provided in the allowed regimes (where $\omega / H \ll 1$ and of course $M_{\mathrm{eff}} / H \ll 1$). We also list the bounds \pref{12bound1}-\pref{12bound3} for completeness. The bound in Table \ref{NMBounds1} require less extreme values of $g$, $\omega/H$ and $M_{\mathrm{eff}} / H$ to be satisfied (than that required by the Markovian limit in \S\ref{sec:MarkovianValidity}).
\begin{table}[h]
  \centering    
     \centerline{\begin{tabular}{ r|c|c| }
 \multicolumn{1}{r}{}
 & \multicolumn{1}{c}{$\frac{\omega}{H} \ll \frac{\kappa}{H} \simeq \frac{M_{\mathrm{eff}}^2}{3H^2} \ll 1$}
 & \multicolumn{1}{c}{$\underset{\ }{ \frac{\kappa}{H} \simeq \frac{M_{\mathrm{eff}}^2}{3H^2} \ll \frac{\omega}{H} \ll 1 }$} \\
\cline{2-3}
$\left| \dfrac{ g^2 \CBD \tDBDp}{\tCBD} \right|  $ & $ \stackrel{\ }{ \underset{\ }{ \frac{27}{2\pi^2(\pi^2+3)}\dfrac{g^2|\log(H\epsilon)| H^4}{M_{\mathrm{eff}}^4} } } $ & $ \frac{3}{2\pi^2(\pi^2+3)}\dfrac{g^2 |\log(H\epsilon)| H^2}{\omega^2} $ \\
\cline{2-3}
$\left| \dfrac{g^2 \CBD \tCBDp}{\tCBD} \right|  $ & $ \stackrel{\ }{ \underset{\ }{ \frac{3(15-\pi^2)}{10(3+\pi^2)} \dfrac{g^2H^3\omega}{M_{\mathrm{eff}}^4} } } $ & $ \frac{15-\pi^2}{30(3+\pi^2)} \dfrac{ g^2 H}{\omega } $ \\
\cline{2-3}
$\left| \dfrac{\omega \tDBDp}{\tCBD} \right|  $ & $ \stackrel{\ }{ \underset{\ }{ \frac{6}{\pi^2+3}\dfrac{ |\log(H\epsilon)| \omega}{ H}   } } $ & $  \frac{6}{\pi^2+3}\dfrac{ |\log(H\epsilon)| \omega}{ H} $ \\
\cline{2-3}
$\left| \dfrac{\omega \tCBDp}{\tCBD} \right|  $ & $ \stackrel{\ }{ \underset{\ }{ \frac{2(15-\pi^2)\pi^2}{15(3+\pi^2)}\dfrac{ \omega^2}{ H^2 }  } } $ & $   \frac{2(15-\pi^2)\pi^2}{15(3+\pi^2)}\dfrac{ \omega^2}{ H^2 }  $ \\
\cline{2-3}
$\left| \dfrac{\kappa \tDBDp}{2 \tCBD} \right|  $ & $ \stackrel{\ }{ \underset{\ }{ \frac{2}{\pi^2 + 3}\dfrac{|\log(H\epsilon)|M_{\mathrm{eff}}^2}{H^2}  } } $ & $ \frac{2}{\pi^2 + 3}\dfrac{|\log(H\epsilon)|M_{\mathrm{eff}}^2}{H^2}  $ \\
\cline{2-3}
$\left| \dfrac{\kappa \tCBDp}{2 \tCBD} \right|  $ & $ \stackrel{\ }{ \underset{\ }{ \frac{(15-\pi^2)\pi^2}{45(3+\pi^2)} \dfrac{M_{\mathrm{eff}}^2\omega}{H^3} } } $ & $ \frac{(15-\pi^2)\pi^2}{45(3+\pi^2)} \dfrac{M_{\mathrm{eff}}^2\omega}{H^3} $ \\
\cline{2-3}
$\left| \dfrac{g^2 \kappa \CBD}{\omega^2} \right|  $ & $ \stackrel{\ }{ \underset{\ }{  \frac{3}{4\pi^2} \dfrac{g^2 H^4}{\omega^2 M_{\mathrm{eff}}^2} } } $ & $ \frac{1}{12\pi^2} \dfrac{g^2 H^2 M_{\mathrm{eff}}^2}{\omega^4 } $ \\
\cline{2-3}
$\left| \dfrac{g^2 \DBD}{\omega} \right|  $ & $ \stackrel{\ }{ \underset{\ }{ \left| \dfrac{g^2 \log(H \epsilon)}{2\pi^2} + \frac{27}{4\pi^2} \dfrac{g^2H^6}{M_{\mathrm{eff}}^6} \right| } } $ & $ \left| \dfrac{g^2 \log(H \epsilon)}{2\pi^2} - \frac{3}{4\pi^2} \dfrac{g^2H^4}{\omega^2 M_{\mathrm{eff}}^2} \right| $ \\
\cline{2-3}
$\left| \dfrac{g^2 \CBD}{\kappa} \right|  $ & $ \stackrel{\ }{ \underset{\ }{ \frac{27}{4\pi^2} \dfrac{g^2 H^6}{M_{\mathrm{eff}}^6} } } $ & $ \frac{3}{4\pi^2} \dfrac{g^2 H^4}{\omega^2 M_{\mathrm{eff}}^2} $ \\
\cline{2-3}
\end{tabular} \quad \quad }
        \caption{Leading-order behaviour of the functions in the validity bounds \pref{NM12bounds1} and \pref{NM12bounds2} (the sizes of all which must be $\ll 1$), using information from Table \ref{Functions2} (and also Table \ref{Functions1}). The bounds \pref{12bound1}-\pref{12bound3} are also included, keeping in mind that $g^2 \CBD / \kappa \simeq 2g^2 \RWo / (\kappa^2 + \omega^2)$ in the regimes quoted. Note that some of the bounds are trivially satisfied in the quoted regimes (for example $| {\kappa \tCBDp}/ {(2 \tCBD)} | \ll 1$).} \label{NMBounds1}
\end{table}

\subsection*{Recovery of the Markovian Limit}

The Markovian solution for $\varrho_{12}(\tau)$, which is valid in the $\omega / H \ll {\kappa}/{H} \simeq {M_{\mathrm{eff}}^2}/{3H^2} \ll 1$ regime only (as derived in \S\ref{sec:MarkovianSection}), is a limit of the more general solution \pref{NMsol12}. The non-Markovian solution is valid for any $H \tau \gg 1$, while the Markovian approximation is valid for times $\kappa \tau \gg 1$. Considering times $\kappa \tau \gg 1$ the exponentials in \pref{NMsol12} with time-scales $1/\kappa$ become negligible ({\it ie.\ }$e^{-\kappa \tau} \simeq 0$), such that 
\begin{eqnarray}
\varrho^{\ssI}_{12}(\tau) &\simeq& \varrho_{12}(0) \left[ 1 + \frac{2 g^2 \RWo}{(\kappa - i \omega)^2} \right] e^{- g^2 \CBD \tau} \label{NMsol12toMark1} \\
& \ & \quad \quad \quad  + \varrho_{12}^{\ast}(0) \left[ \frac{g^2 \DBD + i g^2 \CBD }{2\omega} (1 - e^{+ 2 i \omega \tau} ) - \frac{2g^2 \RWo}{\kappa^2 + \omega^2} \right] e^{- g^2 \CBD \tau} \notag \ .
\end{eqnarray}
Furthermore when the Markovian validity relations (given in Table \ref{MarkovianTable1}) are satisfied\footnote{This follows for example by noting $\omega/\kappa \lesssim \mathcal{O}(g^2)$ when the Markovian condition $|\omega \DBDp / \CBD|\ll 1$ is satisfied. In this case $\frac{2g^2 \RWo}{\kappa^2 + \omega^2} \simeq \frac{g^2 \CBD}{\kappa} = \frac{g^2 \CBD}{\omega} \times \frac{\omega}{\kappa} \simeq 0$ within the approximations used.} then the combination $\frac{2g^2 \RWo}{\kappa^2 + \omega^2}$ is negligibly small. Using this fact it follows\footnote{In the $\omega \ll \kappa$ regime, the factor $\frac{2g^2\RWo}{(\kappa - i \omega)^2} = \frac{2g^2 \RWo}{\kappa^2 + \omega^2} [ \frac{\kappa^2 - \omega^2}{\kappa^2 + \omega^2} + i \frac{\kappa \omega}{\kappa^2 + \omega^2} ]$ is negligible when $ \frac{2g^2 \RWo}{\kappa^2 + \omega^2}$ is negligible.} that
\begin{eqnarray}
\varrho^{\ssI}_{12}(\tau) &\simeq& \varrho_{12}(0) e^{- g^2 \CBD \tau}  + \varrho_{12}^{\ast}(0) \frac{g^2 \DBD + i g^2 \CBD }{2\omega} (1 - e^{+ 2 i \omega \tau} ) e^{- g^2 \CBD \tau} \notag
\end{eqnarray}
which is precisely the solution for the Markovian off-diagonal equation (see \pref{approx12int} in the interaction-picture). 

\subsection{The non-Markovian diagonal equation}
\label{sec:NMDiagonal}

The solution to the non-Markovian diagonal equation \pref{rho111NM4} follows through in much the same manner, so we provide a more concise overview of its derivation. As in \S\ref{sec:NMOffDiagonal}, the equation can be converted to a system of first-order linear differential equations. Defining the vector of variables $\mathbf{y}(\tau) := \big( y_{1}(\tau), y_{2}(\tau), y_{3}(\tau), y_{4}(\tau), y_{5}(\tau) \big)$ where
\begin{eqnarray}
y_{1}(\tau) & := & \varrho_{11}^{\ssI}(\tau)\\
y_{2}(\tau) & := & \int_0^\tau \exd s\ e^{-\kappa s} \cos(\omega s) \varrho^{\ssI}_{11}(\tau - s) \\
y_{3}(\tau) & := & \int_0^\tau \exd s\ e^{-\kappa s} \sin(\omega s) \varrho^{\ssI}_{11}(\tau - s) \\
y_{4}(\tau) & := & e^{-\kappa \tau} \cos(\omega \tau) \\
y_{5}(\tau) & := & e^{-\kappa \tau} \sin(\omega \tau)\ ,
\end{eqnarray}
and differentiating these functions makes the system of differential equations close again to
\begin{eqnarray}
\dot{\mathbf{y}}(\tau) & = & \mathbb{D} \mathbf{y}(\tau) + \mathbf{b} \label{vecy}
\end{eqnarray}
where
\begin{eqnarray}
\mathbb{D} := \left[ \begin{matrix} - 2 g^2 \tCBD & - 4 g^2 \RWo & 0 & - 2 g^2 \frac{\RWo \kappa + \mathrm{Im}[\mathcal{W}_0] \omega}{\kappa^2 + \omega^2} & - 2 g^2 \frac{\RWo \omega - \mathrm{Im}[\mathcal{W}_0] \kappa}{\kappa^2 + \omega^2} \\
1 & - \kappa & - \omega & 0 & 0 \\
0 & \omega & - \kappa & 0 & 0 \\ 
0 & 0 & 0 & -\kappa & - \omega \\
0 & 0 & 0 & \omega & -\kappa \end{matrix} \right] \  \\
\mathrm{and} \quad \mathbf{b} := \left[ \begin{matrix} g^2 \RBD \\ 0 \\ 0 \\ 0 \\ 0  \end{matrix} \right] \; , \quad \mathrm{subject\ to\ initial\ condition\ } \quad \mathbf{y}(0) = \left[ \begin{matrix} \varrho_{11}(0) \\ 0 \\ 0 \\ 1 \\ 0  \end{matrix} \right] \ . \quad \quad \ 
\end{eqnarray}
The determinant $\det\mathbb{D} = -2g^2 \CBD (\kappa^2 + \omega^2)^2 \neq 0$ implies that $\mathbb{D}$ is invertible. The unique steady-state solution is then given by $\mathbf{y}^{\mathrm{static}} =  - \mathbb{D}^{-1} \mathbf{b}$. Using $(\mathbb{D}^{-1})_{11} = - 1 / (2g^2\CBD)$ means that the steady-state solution for $\varrho_{11}^{\mathrm{static}} = y_{1}^{\mathrm{static}}$ is
\begin{eqnarray}
\varrho_{11}^{\mathrm{static}} & = & - (\mathbb{D}^{-1})_{11} \; g^2 \RBD \ = \ \frac{\RBD}{2\CBD} \ = \ \frac{1}{e^{2\pi \omega / H} + 1}
\end{eqnarray}
reproducing the expected thermal result. The general solution to \pref{vecy} is
\begin{eqnarray}
\mathbf{y}(\tau) = e^{\tau \mathbb{D}} \left( \mathbf{y}(0) + \mathbb{D}^{-1} \mathbf{b} \right) - \mathbb{D}^{-1} \mathbf{b}
\end{eqnarray}
which shows that the matrix exponential governs the approach to the thermal steady state. Again, the eigenvalues $\lambda^{\mathbb{D}}$ of the matrix $\mathbb{D}$ govern the dynamics of the matrix exponential, which are the roots of the characteristic equation
\begin{eqnarray}
0 & = & \left( (\lambda^{\mathbb{D}})^2  + 2 \kappa \lambda^{\mathbb{D}} + \kappa^2 + \omega^2 \right) \bigg( (\lambda^{\mathbb{D}})^3 + 2 \kappa \left[ 1 + \frac{g^2 \tCBD}{\kappa} \right]  (\lambda^{\mathbb{D}})^2 \label{charD} \\
& \ & \quad \quad \quad \quad \quad \quad + ( \kappa^2 + \omega^2 ) \left[ 1 + \frac{4 g^2 \kappa \CBD}{\kappa^2 + \omega^2}  - \frac{4 g^2 \RWo (\kappa^2 - \omega^2 )}{(\kappa^2 + \omega^2)^2} \right] \lambda^{\mathbb{D}}  + 2 g^2 \CBD ( \kappa^2 + \omega^2 )   \bigg) \ .\notag 
\end{eqnarray}
As a perturbation in $g\ll 1$ the above yields eigenvalues, 
\begin{eqnarray}
\lambda_{1}^{\mathbb{D}} &\simeq & -2 g^2 \CBD \left[ 1 + \ldots \right] \ , \notag \\
\lambda_{2}^{\mathbb{D}} & \simeq & - \kappa \left[ 1  - \frac{2 g^2 \RWo}{\kappa^2  + \omega^2 } + \ldots \right] - i \omega \left[ 1  + \frac{2 g^2 \RWo}{\kappa^2  + \omega^2 } + \ldots \right] \notag \ , \\
\lambda_{3}^{\mathbb{D}} & = & \lambda_{2}^{\mathbb{D}\ast}\ , \label{Deigenvalues} \\
\lambda_{4}^{\mathbb{D}} & = & -\kappa - i \omega \ ,  \notag \\
\lambda_{5}^{\mathbb{D}} & = & -\kappa + i \omega \ , \notag
\end{eqnarray}
where the last two eigenvalues are exact. In performing the above expansion the bounds \pref{12bound1} and \pref{12bound3}, repeated for convenience, 
\begin{eqnarray}
\frac{g^2 \kappa \CBD}{\omega^2} & \ll & 1 \ , \label{11bound1} \\
\frac{g^2 \CBD}{\kappa} \simeq \frac{2g^2\RWo}{\kappa^2 + \omega^2} & \ll & 1 \ , \label{11bound2}  
\end{eqnarray}
are sufficient to control the expansion of the eigenvalues \pref{Deigenvalues}. The reason for this is that\footnote{In \pref{boundproof1} the function $0 < 1/(x^2 + 1) \leq 1$ appears, and in \pref{boundproof2} the function $-1 \leq {(x^2 - 1)}/{(x^2 + 1)} < 1$ appears (with $x \to \kappa/\omega$). Since these $\mathcal{O}(1)$ functions come multiplied here with the small dimensionless variables shown, these terms are bounded.}
\begin{eqnarray}
\frac{4g^2\kappa\CBD}{\kappa^2 + \omega^2} & = & 4 \times \frac{g^2 \kappa \CBD}{\omega^2} \times \frac{1}{(\kappa/\omega)^2 + 1} \ \ll \ 1  \label{boundproof1} \\
\frac{4 g^2 \RWo (\kappa^2 - \omega^2 )}{(\kappa^2 + \omega^2)^2} & = & 2 \times \frac{2 g^2 \RWo}{\kappa^2 + \omega^2}  \times \frac{(\kappa/\omega)^2 - 1}{(\kappa/\omega)^2 + 1}  \ \ll \ 1  \label{boundproof2} 
\end{eqnarray}
which bounds the $\mathcal{O}(g^2)$ terms in the second line of the characteristic equation \pref{charD}. As before, $g^2 \tCBD/\kappa \ll 1$ on account of \pref{11bound2}.

The solution relaxes towards the thermal steady state here too with time-scales $\xi_{\ssM}/2$ and $\simeq 1/\kappa$. A tedious calculation results in the perturbative solution for $x_1(\tau) = \varrho^{\ssI}_{11}(\tau)$ where
\begin{eqnarray}
\varrho^{\ssI}_{11}(\tau) & \simeq & \frac{1}{e^{2 \pi \omega/H} + 1} + \left[ \varrho_{11}(0) - \frac{1}{e^{2\pi\omega/H} + 1} - \frac{2g^2 \RWo}{\kappa^2 + \omega^2} \left( 1 - \frac{2(\kappa^2 - \omega^2)}{\kappa^2 + \omega^2} \right) \right] e^{ - 2 g^2 \CBD \tau} \quad \quad \quad \quad \quad \label{NMsol11} \\
& \  & \quad \quad \quad \quad + \frac{2g^2 \RWo}{\kappa^2 + \omega^2} \bigg[ \left( 1 - \frac{2(\kappa^2 - \omega^2)\varrho_{11}(0)}{\kappa^2 + \omega^2} \right) \cos\left( \left[ 1 + \sfrac{2g^2 \RWo}{\kappa^2 + \omega^2} \right] \omega \tau \right) \notag   \\
& \ & \quad \quad \quad \quad \quad \quad \quad \quad  - \left( \frac{\mathrm{Im}[\mathcal{W}_0]}{\RWo} - \frac{4\kappa \omega \varrho_{11}(0)}{\kappa^2 + \omega^2} \right) \sin\left( \left[ 1 + \sfrac{2g^2 \RWo}{\kappa^2 + \omega^2} \right] \omega \tau \right) \bigg] e^{- \left[  1 - \tfrac{2 g^2 \RWo }{\kappa^2 + \omega^2} \right] \kappa \tau} \notag 
\end{eqnarray}
which is real-valued and again valid for $H\tau \gg 1$.

The validity conditions are again derived with an exponential $\sim \exp\left(\left[ -1/\xi + i \Phi \right]\tau\right)$ ansatz, which leads to linear combinations
\begin{eqnarray}
\varrho_{11}^{\ssI}(\tau - s) \simeq \varrho_{11}^{\ssI}(\tau) - s \dot{\varrho}_{11}(\tau) + \ldots & \to & \tCBD - \tDBDp \left( - \sfrac{1}{\xi} + i \Phi \right) + \ldots \ .
\end{eqnarray}
in the Nakajima-Zwanzig equation. Bounding the derivatives in the Taylor series here only requires $|\tCBD| \gg |\tDBDp| \sqrt{ 1/\xi^2 + \Phi^2 }$ with $(\xi, \Phi) \in \{ \; (\xi_{\ssM}/2,0), \; ( 1/\kappa, \pm \omega ) \}$ and so this leads to the validity relations
\begin{eqnarray}
\left| \frac{2g^2 \tDBD \CBD}{\tCBD} \right|\ \ll \ 1 \quad \quad \mathrm{and} \quad \quad \left| \frac{2g^2 \tDBD \sqrt{ \kappa^2 + \omega^2 }}{\tCBD} \right|\ \ll \ 1
\end{eqnarray}
The bounds in Table \ref{NMBounds1} (that were set when solving the off-diagonal non-Markovian solution) are sufficient to satisfy the above bounds also. 

Finally, as before it is also of no surprise that the above solution simplifies to the Markovian solution 
\begin{eqnarray}
\varrho_{11}^{\ssI}(\tau) & \to & \frac{1}{e^{2 \pi \omega/H} + 1} + \left[ \varrho_{11}(0) - \frac{1}{e^{2\pi\omega/H} + 1} \right] e^{- 2 g^2 \CBD \tau}
\end{eqnarray}
when the limit $\kappa \tau \gg 1$ is taken, and factors of $g^2 \CBD/\kappa \simeq {2 g^2 \RWo}/{(\kappa^2 + \omega^2)}$ are neglected (as demanded in the $\omega \ll \kappa$ Markovian regime).

\subsection{Interacting field theories and secular growth}
\label{sec:SecularGrowth}

Issues of secular growth can be explored as in our study of the accelerated qubit in Minkowski space \cite{qubitpaper1,Burgess:2018sou} by adding a $\lambda \phi^4$ interaction to the massless theory, which we briefly discuss here. Including an self-interaction
\begin{eqnarray}
H_{\lambda} & = & \frac{\lambda}{4!} \int_{\Sigma_{t}} d^{3}\bx\ \phi^4 \otimes \boldsymbol{I}
\end{eqnarray}
to the Hamiltonian $H$ introduces secularly growing terms \cite{Nayfeh,Tanaka:1975fn,Chen:1994zza,Chen:1995ena,Bender:1996je,Berges:2004yj,Urakawa:2009my,Ford:1984hs,Dolgov:1994cq,Boyanovsky:2003ui,Tsamis:2005hd,Enqvist:2008kt,Bartolo:2007ti,Giddings:2010nc,Byrnes:2010yc,Gerstenlauer:2011ti,Giddings:2011zd} that again can be resummed by using a coupling-dependent effective mass, which for nominally massless particles has size \cite{Starobinsky:1994bd,Petri:2008ig,Riotto:2008mv,Burgess:2009bs,Burgess:2010dd}
\begin{eqnarray}
M_{\mathrm{eff}}^2 \ \to \ M_{\lambda}^2 & = & \frac{\sqrt{3\lambda}}{4\pi} \; H^2\,,
\end{eqnarray}
which is very small in that $0 < {M_{\lambda}}/{H} \ll 1$, and shows the characteristic non-analytic dependence on $\lambda$ expected from a resummation.

In the interacting theory the effective mass $M_{\lambda}$ therefore lies in precisely in the regime of parameter space associated with critical slowing down. At lowest-order in the expansion of the kernel in the Nakajima-Zwanzig equation (see \cite{qubitpaper1}), the equations of motion are unaltered by the introduction of $H_{\lambda}$. For this reason, the entire earlier analysis applies here with the mass replaced with $M_{\lambda}$, and in particular the non-Markovian analysis in the critical slowing down regime. The qubit relaxes with time-scales
\begin{eqnarray}
\frac{1}{\kappa} \to \frac{3H}{M_{\lambda}^2} \simeq \frac{12\pi}{\sqrt{3\lambda} H} \quad \quad \mathrm{and} \quad \quad \xi_{\ssM} = \frac{1}{g^2 \CBD} \simeq \begin{cases} \ \frac{\lambda}{12 g^2 H} \quad \quad \quad , \  \tfrac{\omega}{H} \ll \frac{\sqrt{3\lambda}}{12\pi} \\  \ \frac{4\pi^2 \omega^2}{g^2 H^3} \quad \quad \quad , \  \frac{\sqrt{3\lambda}}{12\pi} \ll \tfrac{\omega}{H} \end{cases}
\end{eqnarray}
where of course the time-scales must satisfy the validity relations in Table \ref{NMBounds1}.

\section{Conclusions}
\label{sec:Conc}

The Markovian approximation provides a simplistic description of the qubit state when the evolution is constrained to be extremely slow compared to the scales over which the Wightman function varies. Of great interest in this work is the critical slowing down regime, where the effective mass $M_{\mathrm{eff}}$ of the scalar field is small compared to the de Sitter Hubble scale $H$. Here the correlation time for the Wightman function is $\sim H / M_{\mathrm{eff}}^2$, which is large compared to the Hubble time $1/H$.

The Markovian regime constrains the relative sizes of the parameters in the problem --- in this case, the qubit gap $\omega$, the qubit-field coupling $g$, as well as $M_{\mathrm{eff}}$ and $H$. As a general statement, bounding derivatives in a Taylor expansion $\boldsymbol{\varrho}^{\ssI}(\tau - s) \simeq \boldsymbol{\varrho}^{\ssI}(\tau ) + \ldots$  ensures a heirarchy $\omega \ll H$ so that large oscillations do not violate the assumption of slow qubit evolution. Precise validity relations for the Markovian regime have been collected in this study, which become stringent to satisfy in the critical slowing down regime where $M_{\mathrm{eff}} \ll H$. Although a Markovian limit can be controlled in the critical slowing down regime, it interestingly constrains a hierarchy $\omega  \ll {M_{\mathrm{eff}}^2}/{(3H)} \ll H$ (where in particular the regime ${M_{\mathrm{eff}}^2}/{(3H)} \ll \omega \ll H$ is never Markovian).

A non-Markovian analysis in the critical slowing down regime is also applied in this work, where the memory effects introduced by the slow decay of the Wightman function are accounted for. This allows for a more detailed solution for the qubit evolution, which also allows access to solutions in the ${M_{\mathrm{eff}}^2}/{(3H)} \ll \omega \ll H$ regime. Sufficient validity relations are derived (which are slightly less stringent to satisfy than those in the Markovian limit), and the Markovian solutions are recovered as a limit of the non-Markovian solutions derived. The basic non-Markovian analysis in this work comes with considerable effort, so seemingly a lesson to be learned in the larger scope of Open EFT methods is that Markovianity is much easier to control than tracking even a small degree of non-Markovian effects.

\section*{Acknowledgements}
This work was partially supported by funds from the Natural Sciences and Engineering Research Council (NSERC) of Canada. Research at the Perimeter Institute is supported in part by the Government of Canada through NSERC and by the Province of Ontario through MRI.

\appendix

\section{Late-time asymptotics for the Bunch-Davies Wightman function}
\label{App:WBDLargeTime}

Here we derive the details of Table \ref{BDasymptotics} from \S\ref{sec:Wightman}. We begin with the analytic continuation of the Gauss hypergeometric function \cite{Erd}
\begin{eqnarray}
_2F_1\left(a, b ; c ; z\right) & = & \frac{\Gamma(b-a)\Gamma(c)}{\Gamma(b)\Gamma(c-a)} (-z)^{-a} \; _2F_1\left( a,  a - c + 1 ; a - b +1 ; \sfrac{1}{z} \right) \label{ErdHyper} \\
&\ & \ \ \ \ \ \ \ \ \ \ + \frac{\Gamma(a-b)\Gamma(c)}{\Gamma(a)\Gamma(c-b)} (-z)^{-b} \; _2F_1\left( b,  b - c + 1 ; b - a +1 ; \sfrac{1}{z} \right) \ , \notag 
\end{eqnarray}
which is valid for $a-b \notin \mathbb{Z}$ and $ z \notin (0,1)$. Using the first term in the series representation of $_2F_1$ we get the $|z| \to \infty$ asymptotics in the case that $a-b\notin \mathbb{Z}$, where 
\begin{eqnarray}
_2F_1\left(a, b ; c ; z\right) & \simeq & \frac{\Gamma(b-a)\Gamma(c)}{\Gamma(b)\Gamma(c-a)} (-z)^{-a} + \frac{\Gamma(a-b)\Gamma(c)}{\Gamma(a)\Gamma(c-b)} (-z)^{-b} \ .
\end{eqnarray}
Using the identity $\Gamma(\frac{3}{2} - \nu)\Gamma(\frac{3}{2}+\nu) = \pi \sec(\pi\nu) (\frac{1}{4} - \nu^2)$ we write the Wightman function (\ref{WightmanBD}) as
\begin{eqnarray}
\mathcal{W}_{\mathrm{BD}}(\tau ) & = & \frac{H^2\Gamma(\frac{3}{2}+ \nu) \Gamma(\frac{3}{2} - \nu)}{16\pi^2} \; _2F_1 \left( \tfrac{3}{2} + \nu, \tfrac{3}{2} - \nu ; 2 ; 1 + \big( \sinh( \tfrac{H\tau}{2} ) - i \tfrac{H\epsilon}{2} \big)^2 \right) \ . \label{WBDgammas}
\end{eqnarray}
Then in the limit $H\tau \gg 1$ and for $2\nu\notin \mathbb{Z}$ the Wightman function (\ref{WightmanBD}) takes the approximate form
\begin{eqnarray}
\mathcal{W}_{\mathrm{BD}}(\tau ) & \simeq & \frac{H^2}{16\pi^2} \left[ \frac{\Gamma(-2\nu)\Gamma(\frac{3}{2} + \nu)}{\Gamma(\frac{1}{2} - \nu)} \left( - z(\tau) \right)^{-\tfrac{3}{2} - \nu} + \frac{\Gamma(2\nu)\Gamma(\frac{3}{2} - \nu)}{\Gamma(\frac{1}{2} + \nu)} \left( - z(\tau) \right)^{-\tfrac{3}{2} + \nu} \right] \ ,
\end{eqnarray}
where $z(\tau) = 1 + \big( \sinh( \tfrac{H\tau}{2} ) - i \tfrac{H\epsilon}{2} \big)^2 \simeq \frac{1}{4} e^{H\tau}$ and we have used $z \Gamma(z)= \Gamma(z+1)$. Note the above does not capture the asymptotics for $\nu \in \{ 0, \tfrac{1}{2}, 1 \}$ (equivalently $M_{\mathrm{eff}}/H \in \{ \tfrac{3}{2}, \sqrt{2}, \tfrac{\sqrt{5}}{2} \}$). Simplifying this, 
\begin{eqnarray}
\mathcal{W}_{\mathrm{BD}}(\tau ) & \simeq & \frac{H^2 }{4\pi^{5/2}} \left[\Gamma(-\nu) \Gamma(\tfrac{3}{2}+\nu) e^{-\left(\tfrac{3}{2}+\nu\right)( H\tau + i \pi ) } + \Gamma(\nu) \Gamma(\tfrac{3}{2}-\nu) e^{-\left(\tfrac{3}{2}-\nu\right)( H\tau + i \pi ) } \right] \label{aBD}
\end{eqnarray}
where we have used the duplication formula $\Gamma(z)\Gamma(z+\tfrac{1}{2}) = 2^{1-2z} \sqrt{\pi} \Gamma(2z)$. For $\nu \in (0,\tfrac{3}{2}) \setminus \{ \tfrac{1}{2} , 1\}$, we only need to keep the rightmost term in (\ref{aBD}) (which is leading-order) giving us
\begin{eqnarray}
\mathcal{W}_{\mathrm{BD}}(\tau) & \simeq & \frac{H^2 }{4\pi^{5/2}} i e^{i \pi \nu}  \Gamma(\nu) \Gamma(\tfrac{3}{2}-\nu) e^{-\left(\tfrac{3}{2}-\nu\right) H\tau } + \mathcal{O}\big( e^{- \left( \tfrac{3}{2} + \nu \right) H \tau} \big) \ . \label{WBDleadingsmall}
\end{eqnarray}

Taking the real part of the above yields
\begin{eqnarray}
\mathrm{Re}[ \mathcal{W}_{\mathrm{BD}}(\tau ) ] \ \simeq \ - \frac{H^2}{4\pi^{5/2}} \sin(\pi\nu) \Gamma(\tfrac{3}{2} - \nu) \Gamma(\nu ) e^{- \left( \tfrac{3}{2} - \nu \right) H \tau} \ + \ \mathcal{O}\big( e^{- \left( \tfrac{3}{2} + \nu \right) H \tau} \big) \ . \label{realWBDreal}
\end{eqnarray} 
A word of caution on the above asymptotic form when the limit $M_{\mathrm{eff}}/H \ll 1$ is also taken: the sub-leading corrections (coming from higher-order terms in the hypergeometric series in \pref{ErdHyper}) are actually $\mathcal{O}( e^{ - \left(5/2- \nu \right) H \tau } ) \sim \mathcal{O}( e^{ - H \tau } )$ as noted in \pref{subleading} (in the limit $M_{\mathrm{eff}}/H \ll 1$ this is a slightly slower falloff than $\mathcal{O}\left( e^{- \left( 3/2 + \nu \right) H \tau} \right) \sim \mathcal{O}\left( e^{- 3 H \tau} \right)$ written in the above).

For the case of $\nu = i \mu$ where $\mu = \sqrt{ M_{\mathrm{eff}}^2/H^2 - {9}/{4}}$ (when $M_{\mathrm{eff}}/H > {3}/{2}$), we write the relevant factors in polar form such that
\begin{eqnarray}
\mathcal{W}_{\mathrm{BD}}(\tau ) & \simeq & \frac{H^2 }{4\pi^{3/2}} i e^{-\tfrac{3}{2} H \tau } \big| \Gamma(-i\mu) \Gamma(\tfrac{3}{2}+i\mu)  \big| \bigg[ e^{-i \mu H\tau + \mu \pi - i \mathrm{Arg} \left[ \Gamma(i\mu) \Gamma(\tfrac{3}{2}-i\mu) \right] } \\
& \ & \ \ \ \ \ \ \ \ \ \ \ \  \ \ \ \ \ \ \ \ \ \ \ \  \ \ \ \ \ \ \ \ \ \ \ \   \ \ \ \ \ \ \ \ \ \ \ \ \ \ \ \ \ \  + e^{i\mu H\tau - \mu \pi + i \mathrm{Arg} \left[ \Gamma(i\mu) \Gamma(\tfrac{3}{2}-i\mu) \right] } \bigg] \ . \notag
\end{eqnarray}
Taking the real part of the above yields the oscillatory 
\begin{eqnarray}
\mathrm{Re}[ \mathcal{W}_{\mathrm{BD}}(\tau ) ] & \simeq & \frac{H^2}{4\pi^{3/2}} \sqrt{ \frac{1+4\mu^2}{\mu} \tanh(\pi\mu)  } \; e^{-\tfrac{3}{2} H \tau} \sin\left( \mu H \tau + \mathrm{Arg}\left[ \Gamma(\tfrac{3}{2} - i \mu) \Gamma(i\mu) \right] \right)  \ . \ \ \ \ 
\end{eqnarray}

The asymptotics for $\nu = \tfrac{1}{2}$ (the conformal case with $M_{\mathrm{eff}}/H = \sqrt{2}$) is trivial given the representation (\ref{conformalWBD}). The conformal case does not require an extra row in Table \ref{BDasymptotics} since it agrees with the limits $\nu \to \tfrac{1}{2}^{\pm}$ of (\ref{realWBDreal}) . For the cases $\nu \in \{ 0,1 \}$ (or equivilently $M_{\mathrm{eff}}/H \in \{ 3/2, \sqrt{5}/{2} \}$) we use the analytic continuation for $b - a =  m \in \mathbb{Z}$, where
\begin{eqnarray}
_2F_1\left(a, a + m ; c ; z\right) & = & \frac{\Gamma(c)  (-z)^{-a-m}}{\Gamma(a+m)\Gamma(c-a)} \sum_{n=0}^{\infty} \frac{\Gamma(a+n+m) \Gamma(1-c+a+n+m)}{\Gamma(a) \Gamma(1-c+a) n! (n+m)!} z^{-n} \left[ \log(-z) + h_{n} \right] \notag \\
& \ & \ \ \ \ \ \ \ \ \ \ \ \ \ \ \ \ \ \ \ \ \ \ \ \ \ \ \ + \frac{\Gamma(c)}{\Gamma(c-a)} (-z)^{-a} \sum_{n=0}^{m-1} \frac{\Gamma(a+n) \Gamma(m-n)}{\Gamma(a)\Gamma(c-a-n) n!} z^{-n} ,
\end{eqnarray}
with $h_{n} \equiv \psi_0(1+m+n) + \psi_0(1+n) - \psi_0(a+m+n) - \psi_0(c-a-m-n)$ and $\psi_0(z) = \frac{\Gamma'(z)}{\Gamma(z)}$ being the digamma function \cite{Erd}. Taking the leading-order terms for $\nu\in \{0,1\}$, it follows 
\begin{eqnarray}
\big(\nu=0\  \mathrm{or}\ M_{\mathrm{eff}} = \tfrac{3}{2}H \big)\ \ \ \ \ \ \ \mathcal{W}_{\mathrm{BD}}(\tau) &\simeq& - \frac{H^2}{8\pi} e^{-\tfrac{3}{2}H\tau}\ \ \ \ \ \ \ \  \label{nu0asymptotic} \\
\big(\nu=1\  \mathrm{or}\ M_{\mathrm{eff}} = \tfrac{\sqrt{5}}{2}H \big)\ \ \ \ \ \mathcal{W}_{\mathrm{BD}}(\tau) &\simeq& - \frac{3H^2}{8\pi} e^{-\tfrac{5}{2}H\tau}\ \ \ \ \ \ \ \ 
\end{eqnarray}

Note that taking $\nu \to 0^{+}$ in the asymptotic form (\ref{realWBDreal}) results in precisely (\ref{nu0asymptotic}), so this case does not require an extra row in Table \ref{BDasymptotics}.

\section{$\epsilon$-dependence of divergences in $\DBD$ and $\DBDp$}
\label{App:DBD}

Here we derive the asymptotics for the divergent function $\DBD$. Using the Wightman function \pref{WBDgammas}, but with a small-{\it distance} regulator $\epsilon$. The integral \pref{DBDdef} for $\DBD$ is explicitly\footnote{We replace $\sinh(\tfrac{as}{2}) - i \tfrac{H\epsilon}{2} \to \sinh( \tfrac{a[s-i\epsilon]}{2} )$ relative to the form in \pref{WBDgammas}.}
\begin{eqnarray}
\DBD  & = & 2 \lim_{\epsilon \to 0^{+}} \int_{0}^{\infty} \exd s\ \sin(\omega s) \; \mathrm{Re}\left[ \sfrac{H^2\Gamma(\frac{3}{2}+ \nu) \Gamma(\frac{3}{2} - \nu) \; _2F_1 \left( \tfrac{3}{2} + \nu, \tfrac{3}{2} - \nu ; 2 ; 1 + \sinh( \tfrac{H( s - i \epsilon )}{2} )^2 \right)}{16\pi^2}  \right] \ . \quad \quad
\end{eqnarray}
We cannot take the limit $\epsilon \to 0^{+}$ here, so we keep $\epsilon$ small (compared to the other scales) but finite. For $s$ approaching the coincident limit, the Wightman function has the behaviour
\begin{eqnarray}
\WBD(s) & \simeq & - \frac{1}{4\pi^2 (s- i \epsilon)^2} \ . \label{smallsWBDcoincident}
\end{eqnarray}
We proceed in the same manner as the analogous calculation in \cite{qubitpaper1} by subtracting and adding \pref{smallsWBDcoincident} in the expression for $\DBD$ giving
\begin{eqnarray}
\DBD  & = & \Delta_\mathrm{BD} ^{(\mathrm{divergent})} + \Delta_\mathrm{BD} ^{(\mathrm{finite})} \label{DeltaSplit}
\end{eqnarray}
where 
\begin{eqnarray}
\Delta_\mathrm{BD} ^{(\mathrm{divergent})} & = & 2 \int_{0}^{\infty} \exd s\ \sin(\omega s) \; \mathrm{Re}\left[ - \frac{1}{4\pi^2 (s- i \epsilon)^2 } \right] \\
\mathrm{and}\quad \quad \Delta_\ssM ^{(\mathrm{finite})} & = & 2 \int_{0}^{\infty} \exd s\ \sin(\omega s) \; \mathrm{Re}\bigg[ \sfrac{H^2\Gamma(\frac{3}{2}+ \nu) \Gamma(\frac{3}{2} - \nu) \; _2F_1 \left( \tfrac{3}{2} + \nu, \tfrac{3}{2} - \nu ; 2 ; 1 + \sinh( \tfrac{H( s - i \epsilon )}{2} )^2 \right)}{16\pi^2} \quad \quad \quad  \label{finiteDBD} \\
&\ & \quad \quad \quad \quad \quad \quad  \quad \quad \quad \quad \quad \quad  \quad \quad \quad \quad \quad \quad  \quad \quad \quad \quad \quad \quad  + \frac{1}{4\pi^2 (s- i \epsilon)^2 } \bigg] \; . \notag
\end{eqnarray}
The divergent part $\Delta_\mathrm{BD}^{(\mathrm{divergent})}$ was computed in \cite{qubitpaper1}, where\footnote{Where $\mathrm{chi}$ and $\mathrm{shi}$ are respectively the hyperbolic cosine and sine integral functions \cite{grad}, defined by
\begin{eqnarray*}
\mathrm{chi}(z) := \gamma + \log(z) + \int_{0}^{z} \exd t \ \frac{\cos(t) - 1}{t} \quad \quad \quad \mathrm{and} \quad \quad \quad \mathrm{shi}(z) := - \frac{\pi}{2} - \int_{0}^{z} \exd t \ \frac{\sin(t)}{t} \ .
\end{eqnarray*}
}
\begin{eqnarray}
\Delta_\mathrm{BD}^{(\mathrm{divergent})} & = & \frac{\omega}{2\pi^2} \bigg[ \cosh(\omega \epsilon) \mathrm{chi}\left( \omega \epsilon \right) - \sinh(\omega \epsilon) \mathrm{shi}\left( \omega \epsilon \right) \bigg] \ \simeq \  \frac{\omega}{2\pi^2} \bigg[ \log( e^{\gamma} \omega \epsilon ) + \mathcal{O}(\omega^2 \epsilon^2) \bigg] \ . \quad
\end{eqnarray}
in the $\omega \epsilon \ll 1$ limit (using the expansions $\mathrm{chi}(z) \simeq \gamma + \log(z) + \mathcal{O}(z^2)$ and $\mathrm{shi}(z) \simeq z + \mathcal{O}(z^3)$ for $0< z \ll 1$).
 In contrast to this, the limit $\epsilon \to 0^{+}$ can be safely taken in the integral $\Delta_\mathrm{BD}^{(\mathrm{finite})}$ where
\begin{eqnarray}
\Delta_\mathrm{BD}^{(\mathrm{finite})} & = & \frac{H}{2\pi^2} \int_{0}^{\infty} \exd s\ \sin(\tfrac{\omega}{H} z) \; \left( \mathrm{Re}\bigg[ \sfrac{\Gamma(\frac{3}{2}+ \nu) \Gamma(\frac{3}{2} - \nu) \; _2F_1 \left( \tfrac{3}{2} + \nu, \tfrac{3}{2} - \nu ; 2 ; \cosh^2({z}/{2}) \right)}{4} \bigg] +\frac{1}{z^2} \right) \ , \quad \quad \label{DBDfinite1}
\end{eqnarray}
where the integration variable has also been scaled to $s \to z = {Hs}$. Since the argument of the hypergeometric function in \pref{DBDfinite1} is greater than unity, we use the analytic continuation \pref{ErdHyper} so that
\begin{eqnarray}
\Delta_\mathrm{BD}^{(\mathrm{finite})} & = & \frac{H}{2\pi^2} \int_{0}^{\infty} \exd s\ \sin(\tfrac{\omega}{H} z) \; \bigg( \mathrm{Re}\bigg[ \sfrac{\Gamma(-\nu) \Gamma(\tfrac{3}{2}+\nu) \left( - 4 \cosh^2(z/2) \right)^{- \frac{3}{2} - \nu}  \; _2F_1\big( \tfrac{3}{2} + \nu, \tfrac{1}{2} + \nu; 1 + 2 \nu ; \mathrm{sech}^2(z/2) \big)}{\sqrt{\pi}} \notag  \\
& \  & \quad \quad \quad \quad  + \sfrac{\Gamma(\nu) \Gamma(\tfrac{3}{2}-\nu) \left( - 4 \cosh^2(z/2) \right)^{- \frac{3}{2} + \nu}  \; _2F_1\big( \tfrac{3}{2} - \nu, \tfrac{1}{2} - \nu; 1 - 2 \nu ; \mathrm{sech}^2(z/2) \big)}{\sqrt{\pi}} + \frac{1}{z^2} \bigg) \ , \label{DBDfinite2}
\end{eqnarray}
which is valid for $2\nu \notin \mathbb{Z}$. Relating the above hypergeometric functions to associated Legendre polynomials of the first kind, through the formula \cite{NIST} (with $x>0$)
\begin{eqnarray}
_2F_1(a,b;2b;x) = \frac{\sqrt{\pi}}{\Gamma(b)} x^{-b+\tfrac{1}{2}} (1-x)^{\tfrac{b-a}{2}-\tfrac{1}{4}} P_{a-b-\tfrac{1}{2}}^{-b+\tfrac{1}{2}}\left( \frac{2-x}{2\sqrt{1-x}} \right) 
\end{eqnarray}
and then computing the real part in \pref{DBDfinite2} yields (despite the condition in \pref{DBDfinite2}, this formula is valid for $2\nu \in \mathbb{Z}$ as well)
\begin{eqnarray}
\Delta_\mathrm{BD}^{(\mathrm{finite})} & = & \frac{H}{2\pi^2} \int_{0}^{\infty} \exd s\ \sin(\tfrac{\omega}{H} z) \; \bigg( - \sqrt{\frac{\pi}{8}} \mathrm{csch}^{\tfrac{3}{2}}(z) \; \Gamma(\tfrac{3}{2} + \nu) P_{\tfrac{1}{2}}^{-\nu}(\mathrm{coth} z) \label{DBDfiniteend} \\
& \ & \quad \quad \quad \quad  \quad \quad \quad \quad  \quad \quad \quad \quad \quad \quad \quad \quad  - \sqrt{\frac{\pi}{8}} \mathrm{csch}^{\tfrac{3}{2}}(z) \; \Gamma(\tfrac{3}{2} - \nu) P_{\tfrac{1}{2}}^{\nu}(\mathrm{coth} z) + \frac{1}{z^2} \bigg) \ . \notag
\end{eqnarray}
We find no way to analytically compute \pref{DBDfiniteend} here, although this form is easier to use for numerical integration with computer algebra systems. Numerical investigation reveal the logarithmic $\epsilon$-dependences (in terms of $H \epsilon\ll 1$ or $\omega \epsilon \ll1$) given in Table \ref{Functions1} (differentiating these forms with respect to $\omega$ leads to the $\epsilon$-divergences for $\DBDp$).

\section{Conditions for the validity of the Markovian limit}
\label{App:Validity}

For a more detailed derivation of the Markovian limit, we direct the reader to \cite{qubitpaper1}. Beginning from the Nakajima-Zwanzig equation, a Taylor series $\boldsymbol{\varrho}^{\ssI}(\tau - s) \simeq \boldsymbol{\varrho}^{\ssI}(\tau) - s\dot{\boldsymbol{\varrho}}^{\ssI}(\tau) + \ldots$ has been applied so that 
\begin{eqnarray}
\frac{\partial \varrho^{\ssI}_{11}}{\partial \tau} & \simeq & g^2 \int_{-\tau}^{\tau} \exd s\ \WBD(s) e^{- i \omega s} - 4 g^2 \int_0^{\tau} \exd s\ \mathrm{Re}[\WBD(s) ] \, \cos(\omega s) \bigg[ \varrho^{\ssI}_{11}(\tau) - s \frac{\partial \varrho^{\ssI}_{11}}{\partial \tau}  + \ldots \bigg] \, , \quad \quad \quad \\
\frac{\partial \varrho^{\ssI}_{12}}{\partial \tau} & \simeq & i g^2 \DBD \varrho^{\ssI}_{12}(\tau) - 2 g^2 \int_0^\tau \exd s \ \mathrm{Re}[\WBD(s)] e^{+ i \omega s} \bigg[ \varrho_{12}^{\ssI}(\tau) - s \frac{\partial \varrho^{\ssI}_{12}}{\partial \tau}  + \ldots \bigg]  \\
&\ & \quad \quad  \quad \quad  \quad \quad  \quad \quad  \quad \quad  + \ 2 g^2  e^{+ 2 i \omega \tau} \int_0^\tau \exd s \ \mathrm{Re}[\WBD(s)] e^{- i \omega s} \bigg[ \varrho_{12}^{\ssI\ast}(\tau) - s \frac{\partial \varrho^{\ssI\ast}_{12}}{\partial \tau}  + \ldots \bigg] \, , \notag 
\end{eqnarray}
where we note that we have already specified the counter-term as $\omega_1 = - \DBD$ as in \S\ref{sec:MarkovianSection} (so that $\omega$ is the physical qubit gap). Since late times $\tau \gg \tau_c$ are here being probed (with $\tau_c$ the correlation time of $\WBD$), the limits on the integral can be approximated by $\simeq \infty$. Using the definitions \pref{RBDdef}, \pref{CBDdef} and \pref{DBDdef}  this means that the above equations are approximately given by 
\begin{eqnarray}
\frac{\partial \varrho^{\ssI}_{11}}{\partial \tau} & \simeq & g^2 \RBD - 2 g^2 \bigg[ \CBD \varrho^{\ssI}_{11}(\tau) - \DBDp \; \frac{\partial \varrho^{\ssI}_{11}}{\partial \tau} + \ldots  \bigg] \label{rho11withDeriv} \\
\frac{\partial \varrho^{\ssI}_{12}}{\partial \tau} & \simeq &  - g^2 \bigg[ \CBD \varrho_{12}^{\ssI}(\tau) + \left( \DBDp - i \CBDp \right) \frac{\partial \varrho^{\ssI}_{12}}{\partial \tau} + \ldots \bigg] \label{rho12withDeriv} \\
&\ & \quad \quad  \quad \quad  \quad \quad + g^2 e^{+ 2 i \omega \tau} \bigg[ ( \CBD - i \DBD) \varrho_{12}^{\ssI}(\tau) + \left( \DBDp + i \CBDp \right) \frac{\partial \varrho^{\ssI}_{12}}{\partial \tau} + \ldots \bigg] \  . \notag
\end{eqnarray}
where $\prime$ denotes an $\omega$-derivative.

From here we note the form of the Markovian solutions in \pref{Solution11Schro} in \pref{approx12Schro}, and switch to the {\it interaction-picture} via \pref{INTpicturereducedstate} where
\begin{eqnarray}
\boldsymbol{\varrho}^\ssI(\tau) \ := \  e^{+ i \mathfrak{h} \tau } \boldsymbol{\varrho}(\tau) e^{- i \mathfrak{h} \tau} \ .
\end{eqnarray}
A short calculation reveals that $\varrho^{\ssI}_{11}(\tau) = \varrho_{11}(\tau)$ and $\varrho_{12}^{\ssI}(\tau) = e^{+ i \omega \tau} \varrho_{12}(\tau)$. This means that the diagonal solution \pref{Solution11Schro} is unchanged in the interaction-picture, while the off-diagonal solution is
\begin{eqnarray}
\varrho_{12}^{\ssI}(\tau) & \simeq & \varrho_{12}(0) e^{ - \tau/ \xi_{\ssM}} + \varrho_{12}^{\ast}(0) \left( \frac{g^2 \DBD}{2\omega} + i \; \frac{g^2 \CBD}{2 \omega} \right) (1 - e^{2 i \omega \tau}) e^{ - \tau / \xi_{\ssM}} \label{approx12int}
\end{eqnarray}
which is of course given in the non-degenerate regime (where $\omega \gg g^2 \sqrt{ \mathcal{C}_{\mathrm{BD}}^2 + \Delta_{\mathrm{BD}}^2 }\; $). To bound the derivatives we use interaction-picture ansatzes of the form
\begin{eqnarray}
\varrho^{\ssI}_{11}(\tau) & \simeq & A_{1} + A_{2} \; e^{ - 2  \tau / \xi_{\ssM} } \label{11ansatz} \\
\varrho^{\ssI}_{12}(\tau) & \simeq & B_{1} \; e^{ - \tau / \xi_{\ssM}} + B_{2} \; e^{- \tau / \xi_{\ssM} + 2 i \omega \tau } \ = \ \sum\nolimits_{j=1,2}\; B_{j}  e^{- \tau / \xi_{\ssM} + i \Phi_{j} \tau } \label{12ansatz}
\end{eqnarray}
using $1/\xi_{\ssM} = g^2 \CBD$ from \pref{MarkovTimescale},and defining the phases $\Phi_{1} = 0$ and $\Phi_2 = 2\omega$ for the off-diagonal ansatz ($A_{j},B_{j} \in \bC$ are placeholders for the time-independent amplitudes in the corresponding solutions). These ansatzes will be used to bound the derivative terms in \pref{rho11withDeriv} and \pref{rho12withDeriv}.

It is simpler to begin with the diagonal equation. Using the ansatz $\varrho_{11}^{\ssI}(\tau) \propto e^{ - 2 \tau / \xi_{\ssM}}$ results in the RHS of \pref{rho11withDeriv} containing terms
\begin{eqnarray}
\frac{\partial \varrho^{\ssI}_{11}}{\partial \tau} & \supset & - 2 g^2 A_{2} \; e^{- \tfrac{2 \tau}{\xi_{\ssM}}} \bigg[ \CBD - \frac{2 \DBDp}{\xi_\ssM} + \ldots  \bigg] \ .
\end{eqnarray}
To neglect the derivatives here such that $\varrho_{11}^{\ssI}(\tau - s) \simeq \varrho_{11}^{\ssI}(\tau)$ therefore means to satisfy the bound
\begin{eqnarray}
|\CBD| \gg \left| \frac{2\DBD}{\xi_{\ssM}}\right| \ . \label{11boundone}
\end{eqnarray}

To derive validity relations for the off-diagonal equation, a totally analogous procedure is followed. Using the earlier ansatz \pref{12ansatz} which is a linear combination of terms $\propto B_{j} \;  e^{- \tau / \xi_{\ssM} + i \Phi_{j} \tau }$, the RHS of \pref{rho12withDeriv} contains terms
\begin{eqnarray}
\frac{\partial \varrho^{\ssI}_{12}}{\partial \tau} & \supset & - g^2 B_{j} \; e^{- \tfrac{\tau}{ \xi_{\ssM} } + i \Phi_{j} \tau } \bigg[ \CBD + \left( \DBDp - i \CBDp \right) \left( - \frac{1}{\xi_{\ssM}} + i \Phi_j \right) + \ldots \bigg] \label{rho12withDerivTwo} \\
&\ & \quad \quad  \quad + g^2 e^{+ 2 i \omega \tau} B_{j}^{\ast} \; e^{- \tfrac{\tau}{ \xi_{\ssM} } - i \Phi_{j} \tau } \bigg[ ( \CBD - i \DBD)  + \left( \DBDp + i \CBDp \right) \left( - \frac{1}{\xi_{\ssM}} - i \Phi_j \right) + \ldots \bigg] \notag \ .
\end{eqnarray}
Dropping the derivatives in the Taylor series therefore means to satisfy the bounds
\begin{eqnarray}
|\CBD| & \gg & \left| \left( \DBDp - i \CBDp \right) \left( - \frac{1}{\xi_{\ssM}} + i \Phi_{j} \right) \right| \ , \label{conditionONE} \\
|\CBD - i \DBD | & \gg & \left| \left( \DBDp + i \CBDp \right) \left( - \frac{1}{\xi_{\ssM}} - i \Phi_j \right) \right| \  .
\end{eqnarray}
Since $|\CBD - i \DBD| \geq |\CBD|$, bounding the first condition automatically bounds the second. In addition to this, $|\CBD|$ is larger than each of the real and imaginary parts of the complex number on the inside of the modulus in \pref{conditionONE}. This implies that
\begin{eqnarray}
|\CBD| \gg \left| \frac{\DBDp}{\xi_{\ssM}} - \CBDp \Phi_j \right|  \quad \quad  \mathrm{and} \quad \quad  |\CBD| \gg \left| \frac{\CBDp}{\xi_{\ssM}} + \DBDp \Phi_j \right| \label{GeneralValidity}
\end{eqnarray}
are sufficient to bound the derivatives and end up with the desired Markovian equation of motion. For $\Phi_1 = 0$, the above bounds imply
\begin{eqnarray}
|\CBD| \gg \left| \frac{\DBDp}{\xi_{\ssM}} \right|  \quad \quad  \mathrm{and} \quad \quad  |\CBD| \gg \left| \frac{\CBDp}{\xi_{\ssM}} \right| \ , \label{conditionTWO}
\end{eqnarray}
the first of which is essentially the same as \pref{11boundone} (up to a factor of 2). These can be put in the more useful form
\begin{eqnarray}
\frac{1}{\xi_{\ssM}} \ll \left| \frac{\CBD}{\DBDp} \right|  \quad \quad  \mathrm{and} \quad \quad  \frac{1}{\xi_{\ssM}} \ll \left| \frac{\CBD}{\CBDp} \right| \ .
\end{eqnarray}
On the other hand, for $\Phi_{2} = 2\omega$ the earlier bounds imply
\begin{eqnarray}
|\CBD| \gg \left| \frac{\DBDp}{\xi_{\ssM}} - 2 \omega \CBDp \right|  \quad \quad  \mathrm{and} \quad \quad  |\CBD| \gg \left| \frac{\CBDp}{\xi_{\ssM}} + 2\omega \DBDp \right| \ \ . \label{conditionTHREE}
\end{eqnarray}
By explicitly using $1 / \xi_{\ssM} = g^2 \CBD$ in the validity relations \pref{conditionTWO} and \pref{conditionTHREE}, these can be put into the simpler forms
\begin{eqnarray}
g^2 |\DBDp| \ll 1\ , \quad  g^2 |\CBDp| \ll 1 \ , \quad  \left|g^2 \DBDp - \frac{2\omega\CBDp}{\CBD} \right| \ll 1 \quad \mathrm{and} \quad \left| g^2 \CBDp + \frac{2\omega\DBDp}{\CBD} \right| \ll 1 \ , \quad \quad
\end{eqnarray}
which are precisely the validition relations given in \pref{validity1} in \S\ref{sec:MarkovianValidity}.

\section{Singularities and ordering of the limits $\omega \to 0$ and $M_{\mathrm{eff}} \to 0$}
\label{App:OrdLim}

There is a subtlety that arises when computing the asymptotic forms for the functions $\CBD$ and $\DBD$ used in the main text. The subtlety arises because these functions diverge in the limit that $M_{\rm eff}$ and $\omega$ both vanish together, and this makes the asymptotic form found in the regime where both $M_{\mathrm{eff}} \ll H$ and $\omega \ll H$ depend somewhat on the order in which these parameters are made small. 

More precisely, using the asymptotic form $\WBD(\tau) \simeq \mathcal{W}_0 e^{-\kappa\tau}$ in the $M_{\mathrm{eff}}/H\ll 1$ regime (see \pref{WBDsmallMeff}) the relevant functions have the following behaviour
\begin{eqnarray}
\CBD \simeq \frac{2\RWo \kappa}{\kappa^2 + \omega^2}  \quad \hbox{and} \quad
\DBD \simeq \frac{2\RWo \omega}{\kappa^2 + \omega^2} \,, 
\end{eqnarray}
and so
\begin{eqnarray}
 \CBDp \simeq  - \frac{4\RWo \kappa \omega}{(\kappa^2 + \omega^2)^2}   \quad \hbox{and} \quad
  \DBDp \simeq  \frac{2\RWo (\kappa^2 - \omega^2)}{(\kappa^2 + \omega^2)^2} \,,
\end{eqnarray}
where $\kappa \simeq M_{\mathrm{eff}}^2 / (3H)$. These functions therefore have different leading-order asymptotic behaviours near  $(\omega, M_{\rm eff} ) = (0,0)$ depending on which path is chosen to explore it in the $ M_{\mathrm{eff}} - \omega$ plane.

Because of this Table \ref{Functions1} focusses on the leading-order behaviour for these functions in the $\omega/H \ll M_{\mathrm{eff}} / H \ll 1$ regime and vice-versa. For the $\omega/H \ll M_{\mathrm{eff}} / H \ll 1$ regime, it turns out that the sub-leading terms in the asymptotic series for $\CBD$, $\DBD$ and their derivatives are $\mathcal{O}(\omega/\kappa)$ which means that in this case we probe the $\omega \ll \kappa \ll 1$ regime (see Table \ref{Functions1}). For the opposite case of $M_{\mathrm{eff}}/H \ll \omega / H \ll 1$  the sub-leading terms are $\mathcal{O}(\kappa/\omega)$ which means we actually probe the $\kappa \ll \omega \ll 1$ regime (also given in Table \ref{Functions1}).

\end{document}